\documentclass[amsmath,amssymb,showpacs,aps,pre,reprint]{revtex4-1}

\usepackage{txfonts,bbm,hyperref}
\usepackage{graphicx}
\usepackage{color,soul}
\usepackage[T1]{fontenc}
\usepackage{textcomp}

\soulregister\cite7
\soulregister\ref7
\soulregister\pageref7

\begin{document}

\title{Errors in energy landscapes measured with particle tracking}
\author{Micha\l{} Bogdan}
\author{Thierry Savin}
\affiliation{Department of Engineering, University of Cambridge, Cambridge CB2 1PZ, United Kingdom}

\date{08/03/2017}

\begin{abstract}
Tracking Brownian particles is often employed to map the energy landscape they explore. Such measurements have been exploited to study many biological processes and interactions in soft materials. Yet, video tracking is irremediably contaminated by localization errors originating from two imaging artifacts: the ``static'' errors come from signal noise, and the ``dynamic'' errors arise from the motion blur due to finite frame acquisition time. We show that these errors result in systematic and non-trivial biases in the measured energy landscapes. We derive a relationship between the true and the measured potential that elucidates, among other aberrations, the presence of false double-well minima in the apparent potentials reported in recent studies. We further assess several canonical trapping and pair-interaction potentials, by using our analytically derived results and Brownian dynamics simulations. In particular, we show that the apparent spring stiffness of harmonic potentials (such as optical traps) is increased by dynamic errors, but decreased by static errors. Our formula allows for the development of efficient corrections schemes, which we also present in this paper.
\end{abstract}

\maketitle

%%%%%%%%%%%%%%%%%%%%%%%%%%%%%%%%%%%%%%%%%%%%%%%%%%%%%%%%%%%%%%%%%%%%%%%%
\section{Introduction}
%%%%%%%%%%%%%%%%%%%%%%%%%%%%%%%%%%%%%%%%%%%%%%%%%%%%%%%%%%%%%%%%%%%%%%%%

Video tracking of Brownian particles is an important technique that serves multiple purposes. It has been used for decades to study biological and soft matter, and has indeed provided valuable information on the microscale dynamics and structures of these systems \cite{Meijering:2012hb,Chenouard:2014kg,Manzo:2015dc}.  With this technique, one can for instance probe live-cell microenvironments \cite{Courty:2006km,ElBeheiry:2015bv}, study the dynamics of individual proteins in natural settings \cite{Simson:1995eh,Mashanov:2007je}, or image the viral invasion of host cells \cite{Brandenburg:2007dt,Godinez:2009cw}. Extracting mechanical properties of individual biological molecules has also been shown to be possible by measuring the thermal fluctuations of cytoskeletal and membrane filaments \cite{Yasuda:1996bt,LeGoff:2002bt,Jin:2007hq,Noding:2012km}, and of DNA \cite{Dorfman:2013di,Engel:2014kq}. Using single molecule tracking, recent studies have also measured the trapping energetic landscapes confining the movements of membrane receptors \cite{Hoze:2012jd,Masson:2014hr}. Brownian particles tracking has also been used extensively for synthetic soft matter physics. Hence, a central application of this technique has been to determine the microrheology, diffusion rates or mechanical properties of complex fluids \cite{Qian:1991vy,Saxton:1997uka,FatinRouge:2004dt,Waigh:2016kh}. It has also been used to measure colloidal interactions of electrostatic \cite{Crocker:1998gp,Crocker:1994ge} or entropic \cite{Crocker:1999fq,Lin:2001gt} origins, and more recently to map the trapping energies of microchannels \cite{Mojarad:2012js,Krishnan:2010fp,Mojarad:2013hu,Pagliara:2013gi}.

Statistical analysis of Brownian particle trajectories is a prerequisite to extracting observables that can be physically interpreted \cite{Lee:2017fb}. The mean-squared displacement (MSD) is often calculated, as a measure of the time- or population-averaged dynamics of the tracked particles. For example, the MSD enables distinguishing between diffusive, driven, sub-diffusive, hopping or trapped motions \cite{Saxton:1997uk}. 

Reconstructing the underlying energy landscape guiding the particles' dynamics is another insightful analysis of Brownian trajectories, which has been used in many of the aforementioned applications \cite{Hoze:2012jd,Masson:2014hr,Crocker:1998gp,Crocker:1994ge,Crocker:1999fq,Lin:2001gt,Mojarad:2012js,Krishnan:2010fp,Mojarad:2013hu,Pagliara:2013gi}. To calculate this landscape, the statistics of the Brownian particles' positions is measured and assumed to obey Boltzmann distribution \cite{Crocker:1994ge,Oddershede:2002gu,Krishnan:2010fp,Jenkins:2015ida}. Note that this analysis requires only localizing particles in each frame of the video, while calculating the MSD involves the additional, and often non-trivial, step of linking the particles' successive positions into trajectories~\cite{Lee:2017fb}.

Video particle tracking, however, suffers from various sources of errors. In particular, artifacts intrinsic to the imaging detectors can contaminate the trajectory measurements, well beyond the statistical uncertainties arising from finite sampling. Several studies have compared the resilience of tracking methods to these errors \cite{Cheezum:2001ga,Chenouard:2014kg}, and new Bayesian techniques notably tend to improve the robustness of the extracted trajectories \cite{Smal:2008dy,Chenouard:2014kg}. Nevertheless, positioning and trajectory linking are irremediably suffering from errors, which have been recognized to propagate to the measured physical observables \cite{Qian:1991vy,Martin:2002ch,Savin:2005ko,Savin:2005ch,Ritchie:2005fj,Wong:2006da,vanderHorst:2010kt,Michalet:2010hd,Berglund:2010ff,Krishnan:2010fp,Jenkins:2015ida,Hoze:2015jo,Calderon:2016ia,Burov:2017bt,Hoze:2017bc}. 

Most detection errors may be classified into two categories: ``static'' and ``dynamic'' \cite{Savin:2008dv}. The ``static error'' typically comes from video signal noise (camera-specific noise, background autofluorescence, etc) and would even affect the localization of an immobile particle \cite{Savin:2008dv,Lee:2017fb}. The ``dynamic error'' is the result of motion blur, due to finite camera exposure time, and occurs when measuring the positions of a moving particle. The propagation of these errors to MSD calculations has been characterized in detail \cite{Savin:2005ch,Savin:2005ko,Berglund:2010ff}. However, no such systematic description exists for their effects on mapping energetic landscapes. Yet, the need for such studies has been emphasized by the recent experimental work of \citet{Krishnan:2010fp}. If inference schemes are a promising approach to extract reliable measures of trapping potentials from noisy data \cite{Turkcan:2012ev,ElBeheiry:2016jg}, dynamic errors have not yet been incorporated in these schemes.

The goal of this paper is to explain how static and dynamic errors affect energetic mapping. We derive analytically a relationship between the true potential landscape and its apparent evaluation when measurements are contaminated with the errors. Our results notably show that static and dynamic errors cause systematic biases and misinterpretations in experimental results. We also explore means for post-measurement corrections of these errors, which would allow experimentalists to revise their existing data. Implications of our work are general for a wide class of trapping and inter-particle potentials. The article is organized as follows. Section~\ref{sec:errors} details the measurement technique and its associated errors. Section~\ref{sec:apparent} presents and discusses the model which quantifies how static and dynamic errors affect the measured potentials. Sections~\ref{sec:methods} and \ref{sec:examples} describe the simulations to support the predictions of our formula for specific and relevant potentials. Finally, section~\ref{sec:correction} discusses a strategy to correct experimental results for static and dynamic errors.

%%%%%%%%%%%%%%%%%%%%%%%%%%%%%%%%%%%%%%%%%%%%%%%%%%%%%%%%%%%%%%%%%%%%%%%%
\section{Static and dynamic errors}\label{sec:errors}
%%%%%%%%%%%%%%%%%%%%%%%%%%%%%%%%%%%%%%%%%%%%%%%%%%%%%%%%%%%%%%%%%%%%%%%%

The relationship between the potential $V$ probed by the trapped particles and the probability density function (pdf) of their positions ${\bf r}=({\rm x}_1,{\rm x}_2,{\rm x}_3)$ is given by the Boltzmann distribution $f_{\bf r}(\boldsymbol{r})\propto e^{-\beta V(\boldsymbol{r})}$, where $\beta=(k_{\rm B} T)^{-1}$ ($k_{\rm B}$: Boltzmann constant; $T$: temperature). In our notation, $f_{\bf r}(\boldsymbol{r})$ is the joint pdf of ${\bf r}$ evaluated at $\boldsymbol{r}=(x_1,x_2,x_3)$, the space coordinates. In principle, Boltzmann distribution allows experimentalists to recover the energetic landscape by measuring the distributions of positions of the trapped Brownian particles using video microscopy. In practice, however, cameras measure a moving average of positions over a shutter time $\sigma$, to which a zero-mean random vector $\boldsymbol{\xi}$ resulting from instrumental noise is added \cite{Yasuda:1996bt,Keller:2001dw,Savin:2005ch,Savin:2005ko}:
    \begin{equation}\label{eq:integral}
    \overline{\bf r}(t)=\frac{1}{\sigma}\int^t_{t-\sigma} {\bf r}(s){\rm d}s+\boldsymbol{\xi}
    \end{equation}
at time $t$, with $\boldsymbol{\xi}$ independent of ${\bf r}$. The time average in Eq.~(\ref{eq:integral}) results in motion blur or ``dynamic errors'', while the added noise produces the ``static errors'' that would occur even when locating an immobile particle \cite{Savin:2005ch}. Most relevant to quantify the static error is the noise covariance matrix, $\boldsymbol{E}=\langle\boldsymbol{\xi}\boldsymbol{\xi}^\intercal \rangle$, where $\boldsymbol{\xi}^\intercal$ is the transpose of $\boldsymbol{\xi}$, and $\langle\cdot\!\cdot\!\cdot\rangle$ is the average. The noise covariance matrix can often be written $\boldsymbol{E}=\varepsilon^2\boldsymbol{I}$ (with $\boldsymbol{I}$ the identity matrix) in 2D particle tracking where the static errors are isotropic in the observation plane \cite{Savin:2005ch}. In that case, $\varepsilon$ is the spatial resolution of the tracking method, and together with the detector exposure $\sigma$, they quantify the two common sources of errors in particle tracking.

We denote the measured pdf of the measured positions given by Eq.~(\ref{eq:integral}) as $f_{\overline{\bf r}}$. Applying $f_{\overline{\bf r}}(\boldsymbol{r})\propto e^{-\beta \overline{V}(\boldsymbol{r})}$ to this ``apparent'' pdf does not measure the correct potential $V$ in which the particles move, but an apparent potential $\overline{V}$ via:
    \begin{equation}\label{eq:averaged}
    \beta\overline{V}(\boldsymbol{r})=-\ln f_{\overline{\bf r}}(\boldsymbol{r})+\text{constant}\,,
    \end{equation}
with an added, arbitrarily chosen constant that, unless stated otherwise, will be ignored in the remaining.

%%%%%
    \begin{figure}
    \includegraphics[width=8.6cm]{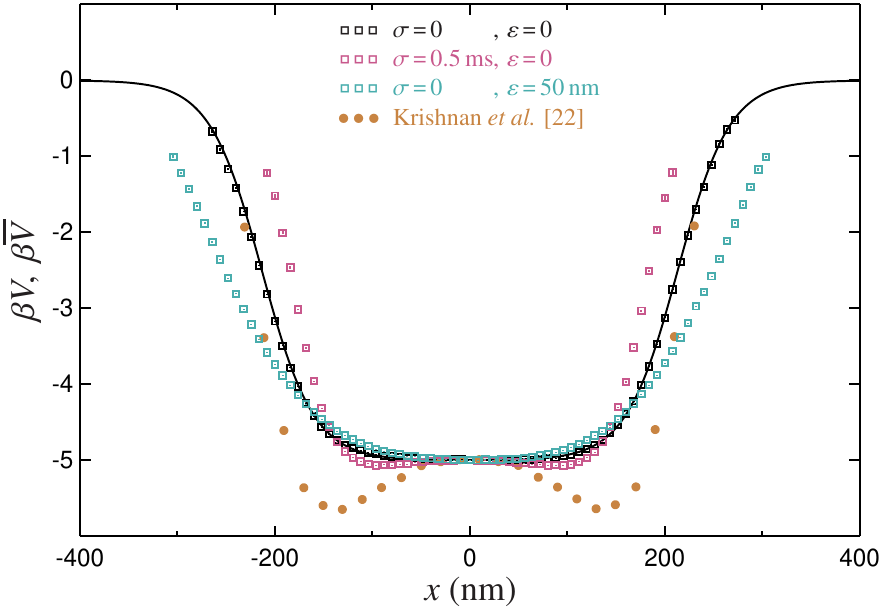}
    \caption{\label{fig1} Effect of static and dynamic errors on a 1D potential mapping. The trapping energetic landscape, $\beta V_\text{ES}(x)= 2.5 [\tanh(3x/a-4)-\tanh(3x/a+4)]$, with $a=160\,\text{nm}$, is shown with the solid line and is chosen to resemble the slice of the 3D electrostatic potential in a microfluidic trap, as measured by \citet{Krishnan:2010fp} using a $100\,\text{nm}$ particle with diffusion coefficient $D=1.8\,\text{\textmu m}^2\text{s}^{-1}$ (fill circles are data reproduced from Fig.~3 of their paper). The open squares are results of our simulations (see section~\ref{sec:methods}). The effect of errors shown in their 3D experimental measurements is more significant than in our 1D simulations, as the dimension may indeed change the magnitude of errors (see section~\ref{sec:2Dpotentials}).}
    \end{figure}
%%%%%

We illustrate in Fig.~\ref{fig1} the effects of errors $\varepsilon$ and $\sigma$ on a representative 1D potential: the exact potential is shown in black, and the measured potentials affected by motion blur or static positional uncertainty are given by the colored squares, which are obtained by  Brownian Dynamics simulations (our algorithm is explained in section~\ref{sec:methods}). While static errors tend to apparently widen the potential (blue squares in Fig.~\ref{fig1}), dynamic errors produce the opposite (red squares). These antagonistic effects were already revealed when studying the propagation of the errors to the mean-squared displacement \cite{Savin:2005ch}. Near the potential's minimum, static errors tend to slightly narrow it, while motion blur gives rise to secondary minima, similar to those observed by \citet{Krishnan:2010fp}, whose results, shown with the orange disk symbols in Fig.~\ref{fig1}, are obtained from 3D tracking measurements (while the simulations presented in Fig.~\ref{fig1} are performed in 1D). The discrepancy in magnitude between the simulations and the measurements is indeed likely to come from this dimensional mismatch (see section~\ref{sec:2Dpotentials}).

%%%%%%%%%%%%%%%%%%%%%%%%%%%%%%%%%%%%%%%%%%%%%%%%%%%%%%%%%%%%%%%%%%%%%%%%
\section{Apparent potential}\label{sec:apparent}
%%%%%%%%%%%%%%%%%%%%%%%%%%%%%%%%%%%%%%%%%%%%%%%%%%%%%%%%%%%%%%%%%%%%%%%%

We derive in appendix~\ref{sec:derivation} the following result for the apparent potential $\overline{V}$:
    \begin{equation}\label{eq:apparent_V}
    \overline{V}=V - \frac{\ln\text{det}(\boldsymbol{U}_{\!\boldsymbol{E},\sigma})}{2\beta}+\frac{\boldsymbol{v}^\intercal\boldsymbol{\Lambda}^{-1}\bigl(\boldsymbol{U}_{\!\boldsymbol{E},\sigma}-\boldsymbol{I}\bigr)\,\boldsymbol{v}}{2\beta D}\,,
    \end{equation}
where $D$ is the diffusion coefficient of the particle, $\boldsymbol{v}=-\beta D \nabla V$ and $\boldsymbol{\Lambda}=\beta D\nabla\nabla^\intercal \!V$ are, respectively, the convective velocity and the local relaxation matrix (with $\nabla$ the nabla vector and $\nabla\nabla^\intercal$ the symmetric Hessian matrix operator; $\boldsymbol{\Lambda}$ has the dimension of an inverse time), and $\boldsymbol{U}_{\!\boldsymbol{E},\sigma}= [\boldsymbol{G}(\sigma \boldsymbol{\Lambda})+\boldsymbol{\Lambda}\boldsymbol{E}/D]^{-1}$ with the matrix function $\boldsymbol{G}(\boldsymbol{X})=2\boldsymbol{X}^{-2}\bigl(\boldsymbol{X}-\boldsymbol{I}+e^{-\boldsymbol{X}}\bigr)$. We use $\text{det}(\cdot\!\cdot\!\cdot)$ to designate the determinant of a matrix. Note that $\boldsymbol{U}_{\boldsymbol{0},0}=\boldsymbol{I}$, indeed leading to $\overline{V}=V$ in the absence of tracking errors.

Eq.~(\ref{eq:apparent_V}) is valid for both shutter times $\sigma$ and static errors small enough to prevent the moving average in Eq.~(\ref{eq:integral}) from blurring third order variation in $V(\boldsymbol{r})$. We write these conditions, conservatively, as:
    \begin{subequations}\label{eq:condition}\begin{align}
    D\sigma\, ||\beta\,\nabla V||+\sqrt{D\sigma} &\ll ||\beta\,\nabla^3\! V||^{-1/3}\,,\\
    ||\boldsymbol{E}||^{1/2} &\ll ||\beta\,\nabla^3\! V||^{-1/3}\,.
    \end{align}\end{subequations}
Here the elements $(\nabla^3\! V)_{ijk}=\frac{\partial^3 V}{\partial x_i\partial x_j\partial x_k}$, and $||\cdot\!\cdot\!\cdot||$ designates the maximum norm, that is, for a position-dependent matrix $\boldsymbol{A}(\boldsymbol{r})$ with elements $A_{ij\dots}(\boldsymbol{r})$,
    \begin{equation}
    ||\boldsymbol{A}(\boldsymbol{r})||=\max_{ij\dots;\,\boldsymbol{r}\in\Omega} |A_{ij\dots}(\boldsymbol{r})|
    \end{equation}
is the maximum absolute value of any elements of the matrix over the observable space domain $\Omega$. The left-hand side of the inequality (\ref{eq:condition}a) represents the typical displacement of the particle during the time $\sigma$, which can be caused by the drift imposed by the trap (first term) and diffusion (second term). We verify in appendix~\ref{sec:validity} that these conditions indeed provide correct limiting values for $\sigma$ and $\varepsilon$ below which Eq.~(\ref{eq:apparent_V}) is valid.

Another requirement for Eq.~(\ref{eq:apparent_V}) to be applicable is $\boldsymbol{U}_{\!\boldsymbol{E},\sigma} \geqslant 0$ (positive definite), which ensures that the logarithmic term is defined. The error matrix $\boldsymbol{E}$ is positive-definite, and we indeed verify that the same holds for $\boldsymbol{G}(\sigma \boldsymbol{\Lambda})$. However, $\boldsymbol{\Lambda}$ does not have this property around local maxima or saddle points. We find that for $\varepsilon>\sqrt{D\sigma}$, $\boldsymbol{U}_{\!\boldsymbol{E},\sigma}$ may not be positive definite at such locations (see section~\ref{sec:1Dpotentials} and Fig.~\ref{fig2}f).

In one dimension, we rename $x_1=x$ and Eq.~(\ref{eq:apparent_V}) is written:
    \begin{equation}\label{eq:apparent_V_1D}
    \overline{V}=V - \frac{\ln u_{\varepsilon,\sigma}}{2\beta}+\frac{v^2\bigl(u_{\varepsilon,\sigma}-1\bigr)}{2\lambda\beta D}\,,
    \end{equation}
with $v =- \beta D \frac{{\rm d} V}{{\rm d}x}=- \beta DV'$, $\lambda = \beta D \frac{{\rm d}^2 V}{{\rm d}x^2}=\beta DV''$, $u_{\varepsilon,\sigma}=[2(\sigma\lambda-1+e^{-\sigma\lambda})/(\sigma\lambda)^2+\lambda\varepsilon^2/D]^{-1}$. The conditions of validity become:
    \begin{subequations}\label{eq:condition_1D}\begin{align}
     D\sigma||\beta V'||+\sqrt{D\sigma} & \ll ||\beta V'''||^{-1/3}\,,\\
    \varepsilon & \ll ||\beta V'''||^{-1/3}\,,
    \end{align}\end{subequations}
with $||f(x)||=\max_{x\in\Omega} |f(x)|$, supplemented with the requirement that $u_{\varepsilon,\sigma}>0$.

One can linearize Eq.~(\ref{eq:apparent_V}) under the more constraining conditions $||\sigma\boldsymbol{\Lambda}||,||\boldsymbol{\Lambda}\boldsymbol{E}/D||\ll1$, to obtain:
    \begin{equation}\label{eq:linear}
    \beta\overline{V}=\beta V +\frac{\text{tr}(\sigma\boldsymbol{\Lambda}\boldsymbol{S}_{\!\boldsymbol{E},\sigma})}{2}-\frac{\sigma\boldsymbol{v}^\intercal\boldsymbol{S}_{\!\boldsymbol{E},\sigma}\boldsymbol{v}}{2D}
    \end{equation}
with $\boldsymbol{S}_{\!\boldsymbol{E},\sigma}=\boldsymbol{E}/(D\sigma)-\boldsymbol{I}/3$ and $\text{tr}(\cdot\!\cdot\!\cdot)$ designating the trace. In particular, Eq.~(\ref{eq:linear}) shows the opposite effects of static and dynamic errors on the apparent potentials, and that these errors can negate each other when $\varepsilon^2=D\sigma/3$, as also observed for the mean-squared displacement of a diffusive particle \cite{Savin:2005ch}.

Typical values of the errors are around $\varepsilon\sim10\,\text{nm}$ and exposure times $\sigma$ in the range of $0.1-1000\,\text{ms}$ for modern CMOS and CCD cameras. The characteristic width $a$ of measurable potentials range from $100\,\text{nm}$ to several microns. The diffusion coefficients of trackable microspheres in a liquid at room temperature are  in the range of $0.1-1\,\text{\textmu m}^2\,\text{s}^{-1}$. Hence, in many instances $\varepsilon\lesssim \sqrt{D\sigma}\lesssim 0.1 a$, and Eq.~(\ref{eq:apparent_V}) should indeed be effective for most experimental settings.

%%%%%%%%%%%%%%%%%%%%%%%%%%%%%%%%%%%%%%%%%%%%%%%%%%%%%%%%%%%%%%%%%%%%%%%%
\section{Methods}\label{sec:methods}
%%%%%%%%%%%%%%%%%%%%%%%%%%%%%%%%%%%%%%%%%%%%%%%%%%%%%%%%%%%%%%%%%%%%%%%%

In the following, we verify the validity of Eq.~(\ref{eq:apparent_V}) by comparing it with Brownian Dynamics (BD) simulations for several examples of potentials. An explicit first-order time-stepping algorithm is used to advance the position ${\bf r}(t)$ of a particle at time $t$: ${\bf r}(t+\delta t)={\bf r}(t)+\dot{\bf r}(t)\delta t$, where $\delta t$ is the time step and $\dot{\bf r}(t)$ satisfies the following equation \cite{Ottinger:1996gx}:
    \begin{equation}\label{eq:BD}
    \dot{\bf r}(t)=-\beta D\,\nabla V\bigl({\bf r}(t)\bigr)+\sqrt{2D/\delta t}\,{\bf w}(t)\,,
    \end{equation}
which assumes the drag on the particle to be Stokesian and neglects any other hydrodynamic interactions. Here, ${\bf w}(t)$ is a Wiener process that satisfies $\langle {\bf w}(t)\rangle=\boldsymbol{0}$ and $\langle {\bf w}(t){\bf w}^\intercal(t')\rangle= \boldsymbol{I}$ if $|t-t'|\le\delta t$, ${\bf 0}$ otherwise.

Each trajectory is $10^9$ time steps long and is then transformed by calculating $\overline{\bf r}(t)=\frac{1}{n+1}\sum^{n}_{k=0}{\bf r}(t-k\delta t)\,+\boldsymbol{\xi}$, where $\sigma=n\delta t$ defines the shutter time, and with $\boldsymbol{\xi}$ a random, normally distributed vector with $\langle\boldsymbol{\xi}\boldsymbol{\xi}^\intercal \rangle=\varepsilon^2\boldsymbol{I}$.

In the remaining, we work with dimensionless quantities, where the unit distance $a$ is the characteristic width of the potential trap (meaning $\beta V(|\boldsymbol{r}|=a)-\beta V(\boldsymbol{0})=1$), the unit energy is $\beta^{-1}$, and the unit time $a^2/D$. In these units, $\delta t$ is chosen to be $5\times10^{-3}$ or less, and $n$ to be $100$ or greater. We further verify, for each simulation, that decreasing $\delta t$ and/or increasing $n$ (while keeping the value $\sigma$ of interest conserved) does not significantly affect the results shown.

%%%%%
    \begin{figure*}
    \includegraphics[width=17.2cm]{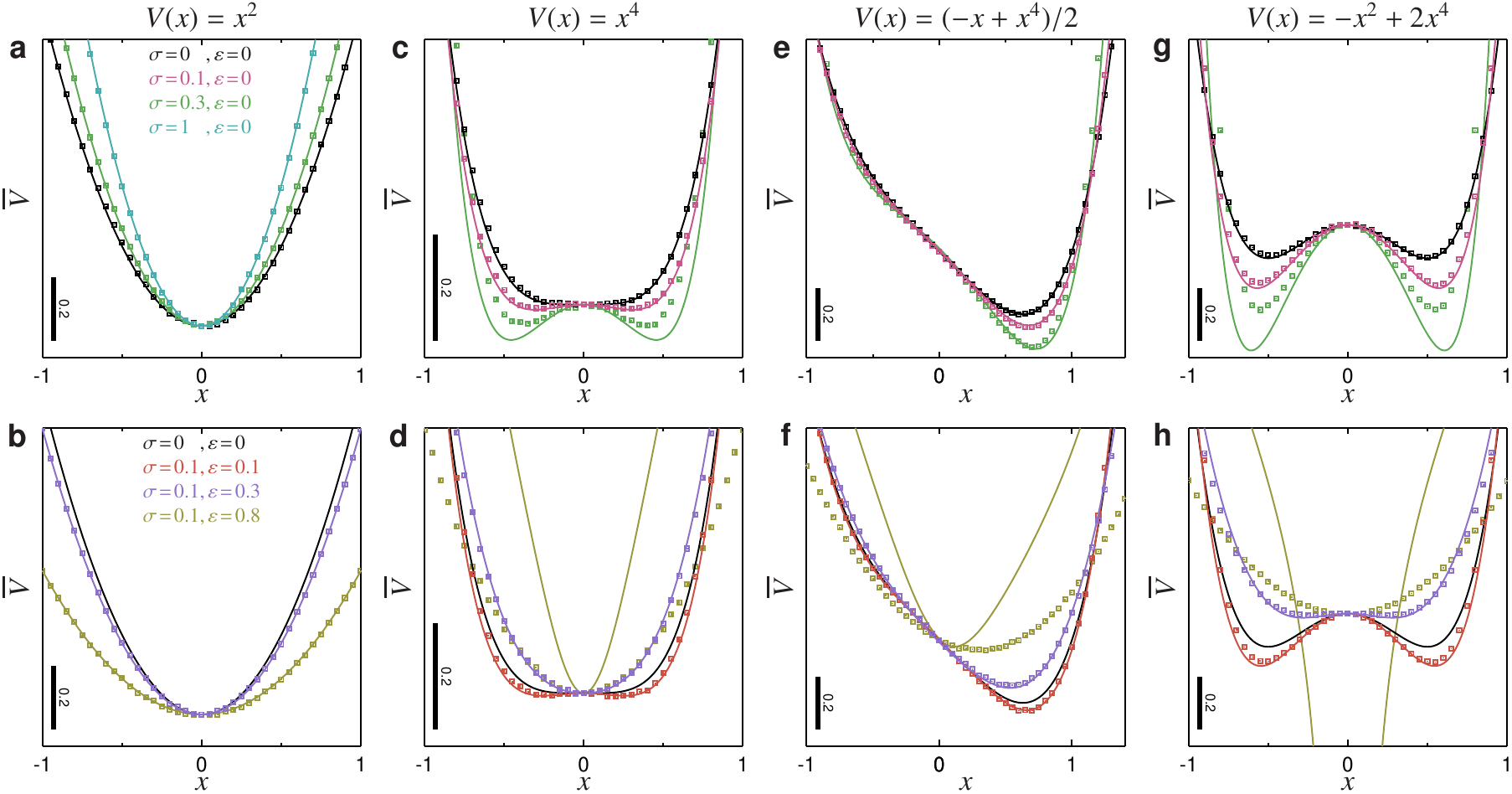}
    \caption{\label{fig2} Comparison of Eq.~(\ref{eq:apparent_V_1D}) (coloured lines on the plots) with BD simulations (symbols) for various $\sigma$ and $\varepsilon$, and under different 1D trapping potentials (black lines): $V(x)=x^2$ (panels a and b), $V(x)=x^4$ (panels c and d), $V(x)=(-x+x^4)/2$ (panels e and f) and $V(x)=-x^2+2x^4$ (panels g and h). The top panels (a, c, e and g) investigate the motion blur with no static errors. The bottom panels (b, d, f and h) concern static errors under a fixed shutter time $\sigma=0.1$. The conditions in Eq.~(\ref{eq:condition_1D}) require $\sigma\ll 0.05$ and $\varepsilon\ll0.3$ for panels c and d,  $\sigma\ll 0.07$ and $\varepsilon\ll0.4$ for panels e and f, $\sigma\ll 0.03$ and $\varepsilon\ll0.3$ for panels g and h.}
    \end{figure*}
%%%%%

A histogram of the positions with a bin size $\le0.05$ is then calculated, from which the apparent potential is extracted via Eq.~(\ref{eq:averaged}). For all examples investigated next, we also perform BD simulations without dynamic and static errors and verify that the correct potential is returned by our algorithm (see Fig.~\ref{fig3}a and the black symbols in Figs.~\ref{fig1}, \ref{fig2}, and \ref{fig4}).

%%%%%%%%%%%%%%%%%%%%%%%%%%%%%%%%%%%%%%%%%%%%%%%%%%%%%%%%%%%%%%%%%%%%%%%%
\section{Examples}\label{sec:examples}
%%%%%%%%%%%%%%%%%%%%%%%%%%%%%%%%%%%%%%%%%%%%%%%%%%%%%%%%%%%%%%%%%%%%%%%%
 
%%%%%%%%%%%%%%%%%%%%%%%%%%%%%%%%%%%%
\subsection{1D potentials}\label{sec:1Dpotentials}
%%%%%%%%%%%%%%%%%%%%%%%%%%%%%%%%%%%%
 
We now use Eq.~(\ref{eq:apparent_V_1D}) to predict the shape of the apparent potential for a few 1D examples presented in Fig.~\ref{fig2}, and compare the results with the BD simulations described in the previous section. In this figure, the lines are obtained from Eq.~(\ref{eq:apparent_V_1D}), while the symbols are obtained from the simulations.

The first potential we consider is $V(x)=x^2$ (Figs.~\ref{fig2}a and \ref{fig2}b), for which Eq.~(\ref{eq:apparent_V_1D}) is exact and indeed matches the simulations for any values of $\varepsilon$ and $\sigma$. For a general harmonic trap with constant $k$, $V(x)=kx^2/2$, the apparent potential can be calculated, using Eq.~(\ref{eq:apparent_V_1D}), as $\overline{V}(x)=\overline{k}x^2/2$ with $\overline{k}=k/u_{\varepsilon,\sigma}$ and for the relaxation rate $\lambda=\beta D k$. Consequently, the apparent mean-squared displacement of a particle trapped in such potential will reach, at long time, a plateau $2/(\beta \overline{k})=2g_\sigma/(\beta k)+2\varepsilon^2$ with $g_\sigma=2(\sigma\lambda-1+e^{-\sigma\lambda})/(\sigma\lambda)^2$, as already shown by \citet{Savin:2005ch}. Our formula in that case also justifies the corrective approach employed by \citet{Mojarad:2012js} to measure the stiffness of their traps.

For the second potential $V(x)=x^4$ (Figs.~\ref{fig2}c and \ref{fig2}d), Eq.~(\ref{eq:apparent_V_1D}) is an approximation that fails for large values of $\sigma$ or $\varepsilon$ (see the green and yellow curves in panels c and d, respectively), when the conditions expressed by Eqs.~(\ref{eq:condition_1D}) are not satisfied. In this case, we observe discrepancies between the predicted apparent potential and the simulations. However, our formula correctly returns the existence of two symmetric minima in the apparent potential, as observed in the simulation results (and similar to the data presented in Fig.~\ref{fig1}) and is accurate for lower (and typically, more experimentally realistic) values of $\sigma$ and $\epsilon$. We also note that near the potential's minimum, the dynamic errors tend to apparently widen the trap, with the static errors producing the reverse. This behavior is indeed the converse of what is seen on the higher parts of the trapping branches of the potential (about $\beta^{-1}$ above its minimum; see Fig.~\ref{fig1}).

We also investigate an asymmetric potential, $V(x)=(-x+x^4)/2$ in Figs.~\ref{fig2}e and \ref{fig2}f, for which Eq.~(\ref{eq:apparent_V_1D}) also returns an effective approximation of the simulation results when $\sigma$ and $\varepsilon$ verify the conditions Eqs.~(\ref{eq:condition_1D}). The potential $V(x)=-x^2+2x^4$, studied in Figs.~\ref{fig2}g and \ref{fig2}h, is symmetric and displays a local maximum at $x=0$ which can be apparently hidden by the static errors (see the purple data, correctly predicted by our formula, in Fig.~\ref{fig2}h). Also in Fig.~\ref{fig2}h, we show an instance where higher values of $\varepsilon$ lead to $u_{\varepsilon,\sigma}<0$ and Eq.~(\ref{eq:apparent_V_1D}) is undefined around a local maximum of $V(x)$ (yellow curves), as explained in section~\ref{sec:apparent}.

We note that overall, Eq.~(\ref{eq:apparent_V_1D}) is returning an effective approximation of the apparent potential $\overline{V}$ unless the static and dynamic errors originate from particularly large values of $\varepsilon$ and $\sqrt{D\sigma}$, respectively, that is, greater than $\sim a/3$.

%%%%%%%%%%%%%%%%%%%%%%%%%%%%%%%%%%%%
\subsection{2D Potentials}\label{sec:2Dpotentials}
%%%%%%%%%%%%%%%%%%%%%%%%%%%%%%%%%%%%

\begin{figure}
    \includegraphics[width=8.6cm]{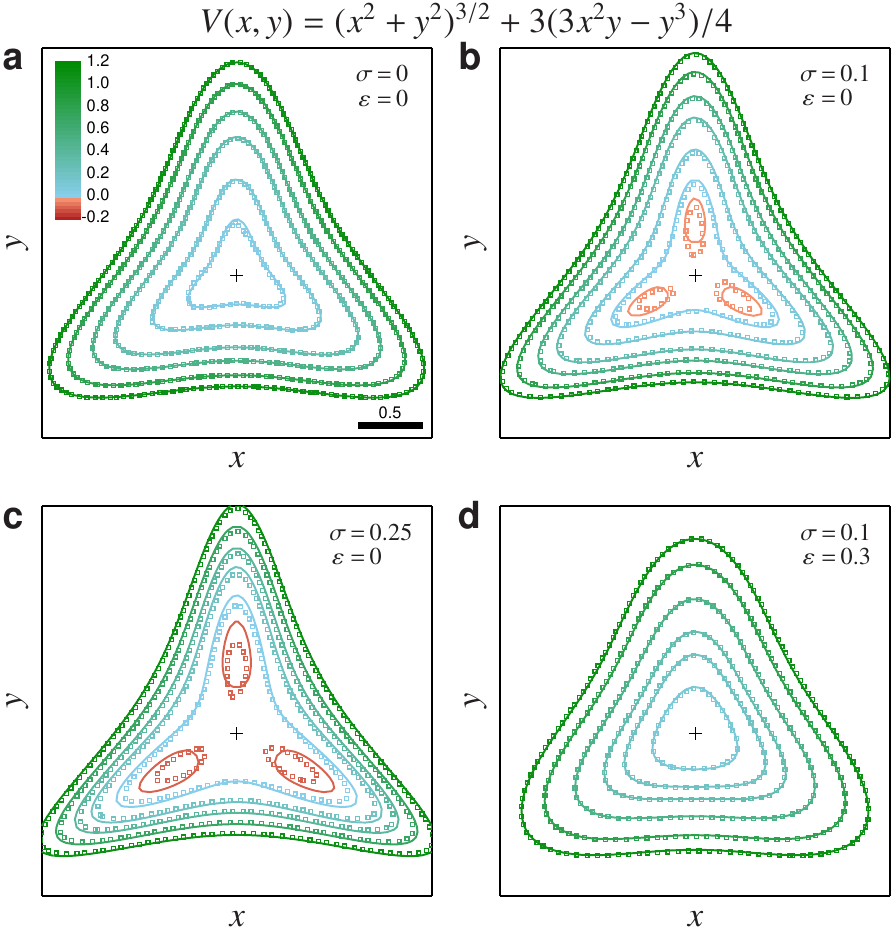}
    \caption{\label{fig3} Comparison of Eq.~(\ref{eq:apparent_V}) with BD simulations for various $\sigma$ and $\varepsilon$, in the two-dimensional trapping potential $V(x,y)=(x^2+y^2)^{3/2}+3(3x^2y-y^3)/4$. The symbols are contours extracted from the simulations, while the lines are their counterparts obtained using Eq.~(\ref{eq:apparent_V}). Panels a-c are for increasing dynamic errors but no static errors, while panel d includes both effects. The cross indicates the point $(0,0)$ and the outermost contour is at $V=1.1$ in all 4 panels. The conditions of Eqs.~(\ref{eq:condition}) require $\sigma\ll 0.1$ and $\varepsilon\ll0.5$ over the observed domain.}
    \end{figure}

We further extend our analysis to 2D potentials and confirm the applicability of Eq.~(\ref{eq:apparent_V}) in that case. In Fig.~\ref{fig3}, we rename $(x_1,x_2)=(x,y)$ and study the potential $V(\boldsymbol{r})=(x^2+y^2)^{3/2}+3(3x^2y-y^3)/4$ (that is, $V(\boldsymbol{r})=r^3\bigl(1+\frac{3}{4}\sin3\theta\bigr)$ in polar coordinates), which traps the particle in a 3-branches star. We show that BD simulations (symbols) are indeed effectively described by Eq.~(\ref{eq:apparent_V}) (lines), even for values of $\sigma$ and $\varepsilon$ close to the limits set by Eqs.~(\ref{eq:condition}).

We further observe that the effects of the dynamic errors share features of the 1D case. Hence, it also produces apparent local minima (see red contours in Figs.~\ref{fig3}b and \ref{fig3}c). But it also modifies the overall shape by sharpening and extending the corners while narrowing the side edges (compare Figs.~\ref{fig3}a, \ref{fig3}b and \ref{fig3}c with increasing $\sigma$ and no static errors). This observation, in particular, does not align with the results reported by \citet{Ritchie:2005fj} in a square confinement (where the particle ``bounces'' on the edge), and indeed highlights the non-trivial effects of motion blur, which depends on the local dynamics of the tracked particle.

%%%%%
%%%%%

The static errors have the opposite effect in the observed range of potential near its minimum, where the corners appear flushed (compare Figs.~\ref{fig3}b and \ref{fig3}d) and the trap narrower. At higher values of the potential, this effect reverses and the potential indeed appears to be widened by the static errors (while, overall, narrowed by the dynamic errors). These considerations may be particularly relevant when studying confined diffusion.

%%%%%%%%%%%%%%%%%%%%%%%%%%%%%%%%%%%%
\subsection{Interaction potential}\label{sec:interaction}
%%%%%%%%%%%%%%%%%%%%%%%%%%%%%%%%%%%%

Eq.~(\ref{eq:apparent_V}) is written for a Brownian particle diffusing in a trapping potential $V$. However, it is also correct for a system of 2 Brownian particles with trajectories ${\bf r}_1(t)$ and ${\bf r}_2(t)$ in a mutual interaction potential $V(|\boldsymbol{r}_1-\boldsymbol{r}_2|)$. One only needs to replace in Eq.~(\ref{eq:apparent_V}) the diffusion constant $D$ with the sum of the diffusion constants of the two particles $D_1+D_2$, and noise covariance matrix with the sum of the individual noise $\boldsymbol{E}_1+\boldsymbol{E}_2$. If the particles are identical and tracked in 1D or 2D, the substitutions are $D\rightarrow 2D$ and $\varepsilon^2\rightarrow2\varepsilon^2$.

This reasoning is valid because Eq.~(\ref{eq:integral}) can also be written identically for the two-particle system, with ${\bf r}={\bf r}_1-{\bf r}_2$ now representing the separation between the interacting particles and the added individual noise vectors $\boldsymbol{\xi}_1$ and $\boldsymbol{\xi}_2$ mutually independent, and because the system's dynamics are now also governed by Eq.~(\ref{eq:BD}), with $D$ replaced with $D_1+D_2$ as obtained by subtracting each Brownian dynamics equation describing ${\bf r}_1(t)$ and ${\bf r}_2(t)$. From there, the derivation of Eq.~(\ref{eq:apparent_V}), as described in appendix~\ref{sec:derivation}, proceeds in an identical manner.

%%%%%
    \begin{figure}
    \includegraphics[width=8.6cm]{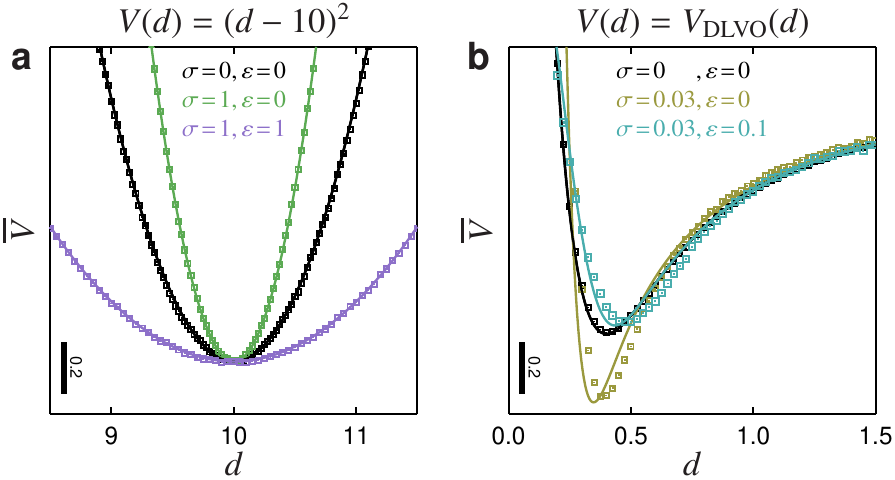}
    \caption{\label{fig4} Comparison of Eq.~(\ref{eq:apparent_V_1D}) with the simulations of two identical Brownian particles interacting via the potential $V(d)=(d-10)^2$ (panel a) and via $V(d)=V_\text{DLVO}(d)=\frac{250e^{-10d}}{d+10}-\frac{50/3}{(d+10)^2}-\frac{50/3}{d(d+20)}-\frac{1}{3}\!\ln\frac{d(d+20)}{(d+10)^2}$ (panel b). In both panels, the symbols are BD results, and the lines are calculated using Eq.~(\ref{eq:apparent_V_1D}) with the substitutions $D\rightarrow 2D$ and $\varepsilon^2\rightarrow2\varepsilon^2$. The conditions in Eqs.~(\ref{eq:condition_1D}) require $\sigma\ll$0.01 and $\varepsilon\ll$0.2 for panel b. }
    \end{figure}
%%%%%

We numerically verify Eq.~(\ref{eq:apparent_V_1D}) for two interaction potentials between identical particles in 1D motion, and the results are presented in Fig.~\ref{fig4}. The first potential models two particles connected by a linear spring with rest length 10, $V(d)=(d-10)^2$ where $d$ is the distance between the particles' surface. For this harmonic potential, Eq.~(\ref{eq:apparent_V_1D}) is exact (see Fig.~\ref{fig4}a) for all values of $\sigma$ and $\varepsilon$. We perform this simulation to verify that the substitutions $D\rightarrow 2D$ and $\varepsilon^2\rightarrow2\varepsilon^2$ are indeed correct.

A relevant interaction in colloidal science is modeled by the Derjaguin-Landau-Verwey-Overbeekthe (DLVO) theory. For a typical system of trackable particles, the potential may be written as \cite{Crocker:1998gp}:
    \begin{equation}\begin{split}\label{eq:DLVO}
   V_\text{DLVO}(d)=	&A_\text{y}\frac{\rho e^{-d/\lambda}}{d+2\rho}-\frac{A_\text{c}}{6}\biggl[\frac{2\rho^2}{(d+2\rho)^2}\\
   						&\qquad+\frac{2\rho^2}{d(d+4\rho)}+\ln\frac{d(d+4\rho)}{(d+2\rho)^2}\biggr]\,,
    \end{split}\end{equation}
with $A_\text{y}=50\beta^{-1}$ (for example, $500\,\text{nm}$ radius particles with $10^{-4}\,\text{C}\,\text{m}^{-2}$ charge density), $A_\text{c}=2\beta^{-1}$ the Hamaker constant for latex particles in water, $\rho=50\lambda$ the particles' radius (for example, in a 1:1 electrolyte with $10^{-3}\,\text{M}$ ionic strength, the Debye length $\lambda=10\,\text{nm}$), and $d+2\rho$ the distance separating the two particles' centers \cite{Israelachvili:2011ug}. To perform the simulations of two Brownian particles interacting with this potential, we set the unit of length to $a=10\lambda$ and the particles are further trapped by a parabolic branch for $d\ge5$, which mimics the effect of the line-scanned optical tweezer used to perform experimental measurements of this kind \cite{Crocker:1999fq}.

The results of our simulations for the DLVO potential are shown in Fig.~\ref{fig4}b, and Eq.~(\ref{eq:apparent_V_1D}) is in reasonable agreement with these data. The effect of dynamic errors is to apparently deepen the interaction potential, and shorten its range. Such systematic differences between true and apparent potentials also occur with interactions of similar profiles, and indeed resemble previously reported mismatches between the experiments and theory \cite{Lin:2001gt,Lau:2002ft}.

%%%%%%%%%%%%%%%%%%%%%%%%%%%%%%%%%%%%%%%%%%%%%%%%%%%%%%%%%%%%%%%%%%%%%%%%
\section{Corrections}\label{sec:correction}
%%%%%%%%%%%%%%%%%%%%%%%%%%%%%%%%%%%%%%%%%%%%%%%%%%%%%%%%%%%%%%%%%%%%%%%%

%%%%%
    \begin{figure}
    \includegraphics[width=8.6cm]{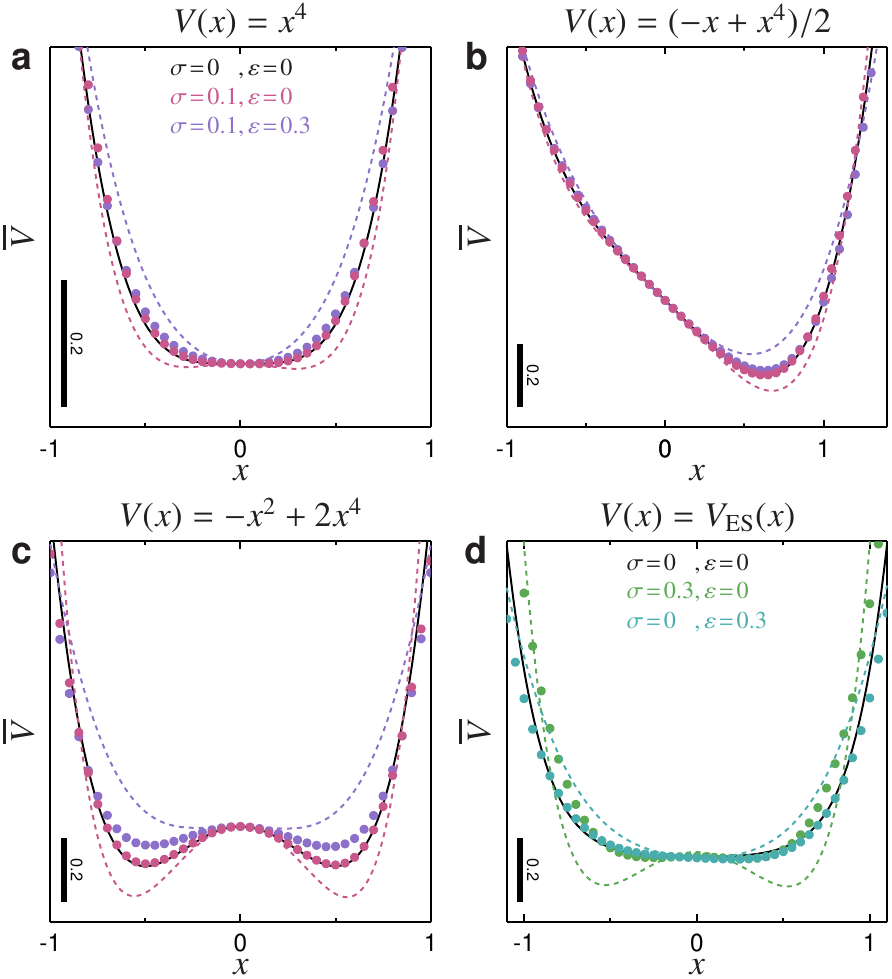}
    \caption{\label{fig5} Corrections of the errors using polynomial fitting. In all panels, the solid lines show the original potentials, while the dashed lines are polynomial fits of the apparent potentials as measured from simulation data affected by the indicated dynamic and static errors. For each set of errors, the symbols show the corrected potentials using polynomial coefficient fitting following Eqs.~(\ref{eq:apparent_V_1D_approx}) and (\ref{eq:poly_V_1D_approx}). Panel a uses data for $V(x)=x^4$ (see Figs.~\ref{fig2}c and \ref{fig2}d), panel b is for $V(x)=(-x+x^4)/2$ (see Figs.~\ref{fig2}e and \ref{fig2}f), panel c for $V(x)=-x^2+2x^4$ (from Figs.~\ref{fig2}g and \ref{fig2}h), and panel d for $V(x)=V_\text{ES}(x)$ as defined in the caption of Fig.~\ref{fig1} with $a=1$.}    
    \end{figure}
%%%%%

In principle, Eq.~(\ref{eq:apparent_V}) is a differential equation that could be solved numerically for $V$ after measuring $\overline{V}$, with a set of boundary conditions (one of which would arbitrarily set the value of $V$ at a particular location). We could not, however, implement a systematic and general solution using common solver packages. 
Instead, we have developed a provisional procedure, which first allows for preliminary assessing if positioning errors are significant in the measurements, and then for obtaining an estimation of the true potential from the apparent potential if the role of these errors is estimated as important.

The measured potential must be first fitted by a power series, using any build-in package for polynomial fitting in the data analysis software. To assess if motion blur can be neglected, one can apply the transformation $V\rightarrow\overline{V}$ described by Eq.~(\ref{eq:apparent_V}) to the fitted apparent potential. If changes are within experimental error bars, no correction needs to be applied. This reasoning is justified by the fact that the transformation described in Eq.~(\ref{eq:apparent_V}) changes the function it is applied to by a comparable factor when applied for the second time, as it does when applied for the first time, as we have numerically verified. 

If applying Eq.~(\ref{eq:apparent_V}) to the apparent potential recovered from data shows changes exceeding experimental error bars, one can estimate the true potential by applying polynomial coefficient fitting of Eq.~(\ref{eq:apparent_V}). For example, if in 1 dimension $||\sigma\lambda||<1$ and $\varepsilon^2<D\sigma$, one can efficiently approximate Eq.~(\ref{eq:apparent_V_1D}) by:
    \begin{equation}\begin{split}\label{eq:apparent_V_1D_approx}
    \beta\overline{V}=\beta V &+ s_{\varepsilon,\sigma}\frac{\sigma\lambda-\sigma\nu^2/D}{2}\\
    &+(1-6s_{\varepsilon,\sigma}^2)\frac{\sigma^2\lambda^2}{24}-(1-12s_{\varepsilon,\sigma}^2)\frac{\sigma^2\lambda\nu^2/D}{24}\\
    &+(1+10s_{\varepsilon,\sigma}-60s_{\varepsilon,\sigma}^3)\frac{\sigma^3\lambda^2\nu^2/D}{120}+c\,,
    \end{split}\end{equation}
with $s_{\varepsilon,\sigma}=
\frac{\varepsilon^2}{(D\sigma)}-
\frac{1}{3}$, as obtained by a second order expansion of $\ln u_{\varepsilon,\sigma}$ and $(u_{\varepsilon,\sigma}-1)/(\sigma\lambda)$ in $\sigma\lambda$ (one order beyond Eq.~(\ref{eq:linear})). In the above equation, $c$ is the constant found in Eq.~(\ref{eq:averaged}). We next write $\beta\overline{V}(x)=\sum_{k=0}^n \overline{c}_k (x/a)^k$ and $\beta V(x)=\sum_{k=0}^n c_k (x/a)^k$ as two polynomial expansions of degree $n$, and where $\{\overline{c}_k\}_{k=0\dots n}$ are the fitting parameters for the measured potential. Upon substituting these expressions into Eq.~(\ref{eq:apparent_V_1D_approx}), and comparing the polynomial coefficients, we obtain a system of equation:
    \begin{equation}\label{eq:poly_V_1D_approx}\begin{split}
    f_0(c,c_0,c_1,c_2)&=\overline{c}_0\\
    f_1(c_1,c_2,c_3)&=\overline{c}_1\\
    \dots & \\
    f_{n-3}(c_1,\dots,c_{n-1})&=\overline{c}_{n-3}\\
    f_{n-2}(c_1,\dots,c_{n})&=\overline{c}_{n-2}\\
    f_{n-1}(c_1,\dots,c_{n})&=\overline{c}_{n-1}\\
    f_n(c_1,\dots,c_{n})&=\overline{c}_n
    \end{split}\end{equation}
where the functions $\{f_k\}_{k=0\dots n}$ can easily be obtained using a symbolic mathematical software. These are $n+1$ equations for the $n+2$ unknowns $c,c_0,c_1,\dots,c_n$, the missing equation being the one that sets $c_0$, which can be assigned arbitrarily by choosing, for example, $V(0)=\overline{V}(0)$ (that is, $c_0=\overline{c}_0$). This well-posed system can then be numerically solved to obtain the coefficient $\{c_k\}_{k=0\dots n}$ of the original potential for the known values of $\sigma$, $\varepsilon$ and $D$.

In Fig.~\ref{fig5}, we apply this method to several of the canonical potentials investigated in this paper. We observe that we can indeed recover the appropriate profiles, notably eliminating the apparent double potential wells (see Fig.~\ref{fig5}a for $\sigma=0.1$ and $\varepsilon=0$, and Fig.~\ref{fig5}d for $\sigma=0.3$ and $\varepsilon=0$), and, on the contrary, restoring lacking features of the true potential that are flushed by the static errors (see Fig.~\ref{fig5}c for $\sigma=0.1$ and $\varepsilon=0.3$).

%\colorbox{red}{The polynomial fits are obtained for power series with} \hl{degree $n$ between 6 and 12, chosen so as to obtain the best match with the original potential. Naturally, this requires a prior knowledge of the probed potential, which is often not available. While our findings prove the validity of the approach, more work is required to offer a systematic and robust numerical method to recover} $V$ from $\overline{V}$.

The polynomial fits are obtained for power series with degree $n$ between 6 and 12, chosen so as to obtain the best match with the original potential. However, a prior knowledge of the probed potential is normally not available. In practice, we anticipate that the best choice of $n$ reflects a compromise between fitting the experimental data as accurately as possible, without capturing features originating from statistical uncertainty over small length scales. A natural criterion for choosing the fitting length scale, and hence $n$, could be based on the terms of Eq.~(\ref{eq:condition_1D}) that sets the validity of Eq.~(\ref{eq:apparent_V_1D}) and that is verified in appendix~\ref{sec:validity}.

We shall deal with this issue in more detail in the course of analyzing published experimental works that could be affected by tracking errors. While our findings prove the validity of the inversion approach, more effort is required to offer a systematic and robust numerical method to recover $V$ from $\overline{V}$.

%%%%%%%%%%%%%%%%%%%%%%%%%%%%%%%%%%%%%%%%%%%%%%%%%%%%%%%%%%%%%%%%%%%%%%%%
\section{Conclusions}
%%%%%%%%%%%%%%%%%%%%%%%%%%%%%%%%%%%%%%%%%%%%%%%%%%%%%%%%%%%%%%%%%%%%%%%%

We have determined the effects of dynamic (resulting from motion blur) and static (resulting from instrumental noise) errors on recovering energy landscapes from measured Brownian particle position distributions. We have shown that these two phenomena lead to non-trivial, systematic biases in the measurements, potentially leading researchers to read out and interpret an incorrect apparent potential. In particular, we have described the phenomenology of these effects in more detail on some canonical trapping potentials: harmonic, double well, asymmetric, in 1D and 2D, as well as interaction potentials. For the harmonic case, the contaminated potential is also harmonic with an apparent stiffness constant that can be exactly calculated.

Estimating if static and dynamic errors significantly skew measurements in a given system can be carried out using our results. Equation~(\ref{eq:apparent_V}) for predicting the apparent potential is accurate for many setups, and easily implemented for a wide class of examples. Inverting it to obtain the true potential from the apparent potential poses a challenge for numerical mathematics, and we also proposed a practical strategy to perform this task. 

We conjecture that the effects of these measurement errors may have been overlooked in some existing experimental works \cite{Lin:2001gt,Lau:2002ft}. Hence, we recommend that the effects of these errors should be assumed one of the possible explanations for unexpected results obtained when using particle tracking methods. Including explicit information about the used shutter times, tracking parameters and noise characterization \cite{Savin:2008dv}, should now become a standard practice in reporting research involving Brownian particle video tracking.

Further research needs to be carried out in this direction. Our study should be followed by a detailed review of existing experimental results. It is also necessary to develop systematic algorithms to invert Eq.~(\ref{eq:apparent_V}) for calculating the true from the apparent potential, $\overline{V}\rightarrow V$. Our current method, explained in section~\ref{sec:correction}, indeed has significant shortcomings. Furthermore, this type of error propagation analysis should also be made for the other observables (e.g. pair or van Hove correlation functions \cite{Valentine:2001fx}, two-point microrheology \cite{Crocker:2000ct}, etc) that are extracted from Brownian particle tracking data.

%%%%%%%%%%%%%%%%%%%%%%%%%%%%%%%%%%%%%%%%%%%%%%%%%%%%%%%%%%%%%%%%%%%%%%%%
\begin{acknowledgments}
The authors thank Dr. Madhavi Krishnan for bringing our attention to this problem, and Drs. Krzysztof and Ma\l{}gorzata Bogdan for fruitful discussions.
\end{acknowledgments}
%%%%%%%%%%%%%%%%%%%%%%%%%%%%%%%%%%%%%%%%%%%%%%%%%%%%%%%%%%%%%%%%%%%%%%%%

%%%%%%%%%%%%%%%%%%%%%%%%%%%%%%%%%%%%%%%%%%%%%%%%%%%%%%%%%%%%%%%%%%%%%%%%
%%%%%%%%%%%%%%%%%%%%%%%%%%%%%%%%%%%%%%%%%%%%%%%%%%%%%%%%%%%%%%%%%%%%%%%%
\section*{Appendix}
\appendix

%%%%%%%%%%%%%%%%%%%%%%%%%%%%%%%%%%%%%%%%%%%%%%%%%%%%%%%%%%%%%%%%%%%%%%%%
\section{Derivation of Eq.~(\ref{eq:apparent_V})}\label{sec:derivation}
%%%%%%%%%%%%%%%%%%%%%%%%%%%%%%%%%%%%%%%%%%%%%%%%%%%%%%%%%%%%%%%%%%%%%%%%

For a single particle in an external potential, we start by writing the moving average of particle positions in Eq.~(\ref{eq:integral}), which represents data collected during a single shutter time, as the limit of a discrete series of $n+1$ successive positions taken by the particle every $\sigma/n$ time units, added to a noise term:
    \begin{equation}\label{eq:discrete}
    \overline{\bf r}(t)=\lim_{n \to \infty}\frac{1}{n+1}\sum^{n}_{k=0}{\bf r}_k(t)\,+\boldsymbol{\xi}
    \end{equation}
where ${\bf r}_k(t)={\bf r}(t-\sigma+k\sigma/n)$, such that ${\bf r}_0(t)={\bf r}(t-\sigma)$ and ${\bf r}_{n}(t)={\bf r}(t)$. The particle obeys the inertialess limit of Langevin equation in an external potential. Consecutive positions in the series forming $\overline{\bf r}(t)$ are located infinitesimally close to each other when $n$ is large. Therefore, the motion between them can be treated via the Brownian dynamics,
    \begin{equation}\label{eq:BD_sig}
    {\bf r}_k={\bf r}_{k-1}-(\sigma/n)\,\beta D\,\nabla V\bigl({\bf r}_{k-1}\bigr)+\sqrt{2D\sigma/n}\,{\bf w}\,,
    \end{equation}
where ${\bf w}$ is a vector realization of a delta-correlated, stationary Gaussian process with zero-mean. Hence, $f_{\bf w}(\boldsymbol{w})= \mathcal{N}(\boldsymbol{w};{\bf 0},\boldsymbol{I})$ and the auto-correlation $\langle {\bf w}(t){\bf w}^\intercal(t')\rangle= \boldsymbol{I}$ if $|t-t'|\le\sigma/n$, ${\bf 0}$ otherwise. We here employ the usual notation for the $d$-dimensional normal distribution with mean vector $\boldsymbol{\mu}$ and covariance matrix $\boldsymbol{\Sigma}$,
    \begin{equation*}
    \mathcal{N}(\boldsymbol{r}\,;\boldsymbol{\mu},\boldsymbol{\Sigma})=\frac{e^{-(\boldsymbol{r}-\boldsymbol{\mu})^\intercal\boldsymbol{\Sigma}^{-1}(\boldsymbol{r}-\boldsymbol{\mu})/2}}{(2\pi)^{d/2}\text{det}(\boldsymbol{\Sigma})^{1/2}}\,.
    \end{equation*}
Each position ${\bf r}_{k-1}$ is now assumed to be in the vicinity of ${\bf r}_0$ so that we may linearize the force $\beta D\nabla V(\boldsymbol{r}')=-\boldsymbol{v}_0+ \boldsymbol{\Lambda}_0(\boldsymbol{r}'-\boldsymbol{r}_0)$, where we have defined:
    \begin{equation*}\begin{split}
    \boldsymbol{v}_0&\equiv\boldsymbol{v}|_{\boldsymbol{r}=\boldsymbol{r}_0} \\
    \boldsymbol{\Lambda}_0&\equiv\boldsymbol{\Lambda}|_{\boldsymbol{r}=\boldsymbol{r}_0}
    \end{split}\end{equation*}
for $\boldsymbol{v}$ and $\boldsymbol{\Lambda}$ as defined in section~\ref{sec:apparent}. The subscript ``$0$'' indicates that these are evaluated at $\boldsymbol{r}_0$. The conditions for the second order expansion to be valid are given by Eqs.~(\ref{eq:condition}).

%%%%%
    \begin{figure}
    \includegraphics[width=8.6cm]{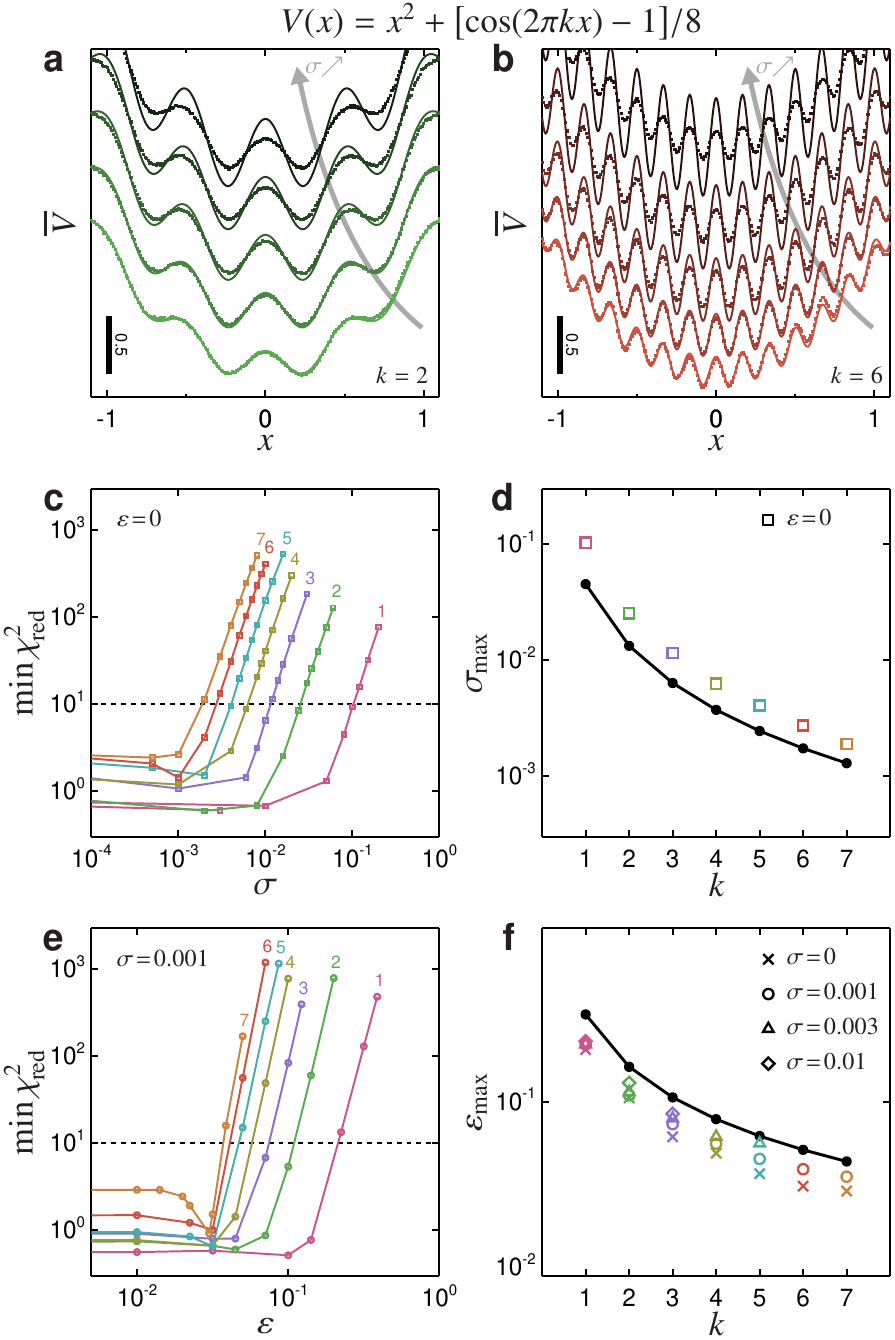}
    \caption{\label{fig6} Assessing the conditions Eqs.~(\ref{eq:condition}) by quantitatively comparing  Eq.~(\ref{eq:apparent_V_1D}) to BD simulations for the trapping potential $V(x)=x^2+[\cos (2\pi k x) -1]/8$, with $k=1\dots7$. Panels a and b show the simulated apparent potentials (symbols) and our approximated expression (lines) for increasing values of $\sigma$ and with $\varepsilon=0$ ($k=2$ in panel a, $k=6$ in b). The discrepancy is quantified by $\text{min}\,\chi^2_{\!\text{red}}$, defined in Eq.~(\ref{eq:redchisqd}), whose variations with $\sigma$ and $k$ are shown in panel c (values of $k$ displayed on each corresponding line). Panel d shows the range $\sigma_{\!\text{max}}$ (below which Eq.~(\ref{eq:apparent_V_1D}) is effective) as a function of $k$, as defined by the threshold $\text{min}\,\chi^2_\text{red}=10$ (symbols), and as obtained by Eq.~(\ref{eq:condition_1D}a) (black line, see text). Panels e and f give the same quantities as panel c and d, respectively, to compare the range of static error $\varepsilon_\text{max}$ evaluated from $\text{min}\,\chi^2_\text{red}=10$ and (\ref{eq:condition_1D}b), for the displayed values of $\sigma<\sigma_\text{max}$.}
    \end{figure}
%%%%%

We now write the conditional pdf $f_{{\bf r}|{\bf r}'}(\boldsymbol{r}|\boldsymbol{r}')=f_{{\bf r},{\bf r}'}(\boldsymbol{r},\boldsymbol{r}')/f_{{\bf r}'}(\boldsymbol{r}')$, in terms of the joint pdf $f_{{\bf r},{\bf r}'}$ and the marginal $f_{{\bf r}'}$. From Eq.~(\ref{eq:BD_sig}) it follows that
    \begin{equation*}\label{eq:recurrence}
    f_{{\bf r}_k|{\bf r}_{k-1}}(\boldsymbol{r}|\boldsymbol{r}')=\mathcal{N}\bigl(\boldsymbol{r}\,;\boldsymbol{A}_n\boldsymbol{r}'+\boldsymbol{b}_n,2D\sigma\boldsymbol{I}/n\bigr)\,,
    \end{equation*}
with $\boldsymbol{A}_n=\boldsymbol{I}-\sigma\boldsymbol{\Lambda}_0/n$ and $\boldsymbol{b}_n=\sigma(\boldsymbol{v}_0+\boldsymbol{\Lambda}_0\boldsymbol{r}_0)/n$. Recursively using $f_{{\bf r}_{k}|{\bf r}_{k-2}}(\boldsymbol{r}|\boldsymbol{r}')=\iiint f_{{\bf r}_k|{\bf r}_{k-1}}(\boldsymbol{r}|\boldsymbol{\rho})f_{{\bf r}_{k-1}|{\bf r}_{k-2}}(\boldsymbol{\rho}|\boldsymbol{r}'){\rm d}^3\rho$ and exploiting properties of Gaussian integrals, we get, for any $k>j\ge 0$: 
    \begin{equation*}
    f_{{\bf r}_k|{\bf r}_j}(\boldsymbol{r}|\boldsymbol{r}')=\mathcal{N}\Biggl(\boldsymbol{r}\,;\boldsymbol{A}_n^{k-j}\boldsymbol{r}'+\!\sum_{i=0}^{k-j-1}\!\boldsymbol{A}_n^i\boldsymbol{b}_n,2D\frac{\sigma}{n}\sum_{i=0}^{k-j-1}\!\boldsymbol{A}_n^{2i}\Biggr)\,.
    \end{equation*}
This equation allows us to calculate $f_{{\bf r}_k|{\bf r}_0}$ and $f_{{\bf r}_k,{\bf r}_j|{\bf r}_0}=f_{{\bf r}_k|{\bf r}_j}f_{{\bf r}_j|{\bf r}_0}$ for any $k>j\ge 1$. All are normal distributions, and so will be $f_{\overline{\bf r}|{\bf r}_0}$. Further using the matrix's geometric series $\sum_{j=0}^{k-1}\boldsymbol{A}_n^j=(\boldsymbol{I}-\boldsymbol{A}_n)^{-1}(\boldsymbol{I}-\boldsymbol{A}_n^k)$, the matrix exponential limit, $\lim_{n\to\infty}(\boldsymbol{I}-\boldsymbol{A}/n)^n=e^{-\boldsymbol{A}}$, and accounting for static errors by adding $\boldsymbol{E}=\langle\boldsymbol{\xi}\boldsymbol{\xi}^\intercal \rangle$ to the covariance matrix of the measured position, we finally obtain
    \begin{equation*}
    f_{\overline{\bf r}|{\bf r}_{0}}(\boldsymbol{r}|\boldsymbol{r}_0)=\mathcal{N}\biggl(\boldsymbol{r}\,;\boldsymbol{r}_0+\frac{\sigma}{2}\boldsymbol{G}_\sigma\boldsymbol{v}_0,D \sigma\boldsymbol{H}_\sigma+\boldsymbol{E}\biggr)\,,
    \end{equation*}
where $\boldsymbol{G}_{\sigma}=\boldsymbol{G}(\sigma \boldsymbol{\Lambda}_0)$ and $\boldsymbol{H}_{\sigma}=\boldsymbol{H}(\sigma \boldsymbol{\Lambda}_0)$ with:
    \begin{equation*}\begin{split}
    \boldsymbol{G}(\boldsymbol{X})&=2\boldsymbol{X}^{-2}\bigl(\boldsymbol{X}-\boldsymbol{I}+e^{-\boldsymbol{X}}\bigr)\,, \\
    \boldsymbol{H}(\boldsymbol{X})&=2\boldsymbol{X}^{-2}-\boldsymbol{X}^{-3}\bigl(3\boldsymbol{I}-e^{-\boldsymbol{X}}\bigr)\bigl(\boldsymbol{I}-e^{-\boldsymbol{X}}\bigr)\,.
    \end{split}\end{equation*}

We now use $f_{\overline{\bf r}}(\boldsymbol{r})=\iiint f_{\overline{\bf r}|{\bf r}_{0}}(\boldsymbol{r}|\boldsymbol{r}_0)f_{{\bf r}_{0}}(\boldsymbol{r}_0){\rm d}^3r_0$ where $f_{{\bf r}_{0}}(\boldsymbol{r}_0)=f_0e^{-\beta V(\boldsymbol{r}_0)}$, with $f_0$ a constant, to calculate the apparent distribution. We may use again the expansion:
    \begin{equation*}\begin{split}
    \beta DV(\boldsymbol{r}_0)&=\beta DV(\boldsymbol{r})-\boldsymbol{v}^\intercal(\boldsymbol{r}_0-\boldsymbol{r})+\frac{1}{2}(\boldsymbol{r}_0-\boldsymbol{r})^\intercal\boldsymbol{\Lambda}\,(\boldsymbol{r}_0-\boldsymbol{r})\,,\\
    \boldsymbol{v}_0&=\boldsymbol{v}-\boldsymbol{\Lambda}\,(\boldsymbol{r}_0-\boldsymbol{r})\,,\\
    \boldsymbol{\Lambda}_0&=\boldsymbol{\Lambda}\,,
    \end{split}\end{equation*}
where $\boldsymbol{v}$ and $\boldsymbol{\Lambda}$ are evaluated at $\boldsymbol{r}$. The resulting integral reads, after the change of variable $\boldsymbol{\rho}=\boldsymbol{r}_0-\boldsymbol{r}$:
    \begin{widetext}
    \begin{equation*}
    f_{\overline{\bf r}}=\frac{f_0e^{-\beta V}}{(2\pi)^{d/2}\text{det}(D \sigma\boldsymbol{H}_\sigma+\boldsymbol{E})^{1/2}}\iiint \exp\biggl\{-\biggl[\boldsymbol{\rho}+\frac{\sigma}{2}\boldsymbol{G}_\sigma\bigl(\boldsymbol{v}-\boldsymbol{\Lambda}\,\boldsymbol{\rho}\bigr)\biggr]^\intercal (D \sigma\boldsymbol{H}_\sigma+\boldsymbol{E})^{-1}\biggl[\boldsymbol{\rho}+\frac{\sigma}{2}\boldsymbol{G}_\sigma\bigl(\boldsymbol{v}-\boldsymbol{\Lambda}\,\boldsymbol{\rho})\biggr]+\frac{\boldsymbol{v}^\intercal\boldsymbol{\rho}}{D}-\frac{\boldsymbol{\rho}^\intercal\boldsymbol{\Lambda}\,\boldsymbol{\rho}}{2D}\biggr\}\,{\rm d}^3\!\rho
    \end{equation*}
    \end{widetext}
After noting the relation $[\boldsymbol{I}-\boldsymbol{X}\boldsymbol{G}(\boldsymbol{X})/2]^2=\boldsymbol{G}(\boldsymbol{X})-\boldsymbol{X}\boldsymbol{H}(\boldsymbol{X})$, we finally obtain:
    \begin{equation*}
    f_{\overline{\bf r}}=\frac{f_0e^{-\beta V}}{\text{det}\bigl(\boldsymbol{G}_{\sigma}+\frac{\boldsymbol{\Lambda}\boldsymbol{E} }{D}\bigr)^{1/2}}\exp\biggl\{-\frac{\boldsymbol{v}^\intercal\boldsymbol{\Lambda}^{-1}\bigl[\bigl(\boldsymbol{G}_{\sigma}+\frac{\boldsymbol{\Lambda}\boldsymbol{E}}{D}\bigr)^{-1}-\boldsymbol{I}\bigr]\,\boldsymbol{v}}{2 D}\biggr\}\,,
    \end{equation*}
from which Eq.~(\ref{eq:apparent_V}) can be readily deduced.

%%%%%%%%%%%%%%%%%%%%%%%%%%%%%%%%%%%%%%%%%%%%%%%%%%%%%%%%%%%%%%%%%%%%%%%%
\section{Conditions of validity}\label{sec:validity}
%%%%%%%%%%%%%%%%%%%%%%%%%%%%%%%%%%%%%%%%%%%%%%%%%%%%%%%%%%%%%%%%%%%%%%%%

We assess here the ranges of $\sigma$ and $\varepsilon$ for which Eq.~(\ref{eq:apparent_V}) can be used. The examples investigated in the main text suggest that the conditions of validity Eqs.~(\ref{eq:condition}) provide appropriate estimates for the maximum values $\sigma_\text{max}$ and $\varepsilon_\text{max}$ below which Eq.~(\ref{eq:apparent_V}) can indeed be used. To assess these limiting values in a systematic manner, we simulated a Brownian particle trapped in the potential $V(x)=x^2+[\cos (2\pi k x) -1]/8$, for $k=1\dots7$, with increasing values of $\sigma$ and of $\varepsilon$. Increasing $k$ for this potential increases the level of details that needs to be resolved by the particle tracking methods (compare Fig.~\ref{fig6}a, where $k=2$, with Fig.~\ref{fig6}b where $k=6$). For this potential, we test our predictions for $\sigma_{\!\text{max}}$ and $\varepsilon_\text{max}$ obtained by equating both sides in each Eq.~(\ref{eq:condition_1D}a) and Eq.~(\ref{eq:condition_1D}b),
    \begin{subequations}\label{eq:maximum}\begin{align}
    (2+\pi k/4)\sigma_{\!\text{max}}+\sigma_{\!\text{max}}^{1/2}&=(\pi k )^{-1}\,,\\
    \varepsilon_\text{max}&=(\pi k )^{-1}\,,
    \end{align}\end{subequations}
respectively. The amplitude $1/8$ of the oscillations around the term $x^2$ in the potential is such that no term may be neglected in Eqs.~(\ref{eq:maximum}).

The simulation results (symbols in Fig.~\ref{fig6}a and \ref{fig6}b) are then compared to the apparent potential $\overline{V}$ predicted by Eq.~(\ref{eq:apparent_V_1D}) (lines in Fig.~\ref{fig6}a and \ref{fig6}b). Specifically, the discrepancy between the simulations and Eq.~(\ref{eq:apparent_V_1D}) is quantified by the reduced chi-squared $\chi^2_{\!\text{red}}$, defined as
    \begin{equation}\label{eq:redchisqd}
    \chi^2_{\!\text{red}}=\frac{1}{N}\sum_{j=1}^N \frac{\Delta\overline{V}_{\!j}^2}{\text{var}\overline{V}_{\!j}}\,.
    \end{equation}
Here, $\{\Delta\overline{V}_{\!j}\}_{j=1\dots N}$ are the differences between the simulations and Eq.~(\ref{eq:apparent_V_1D}) at the $N$ locations output by the simulations, and $\{\text{var}\overline{V}_j\}_{j=1\dots N}$ are the variances of the simulated data at these locations. The arbitrary constant in Eq.~(\ref{eq:averaged}) is chosen beforehand to minimize $\chi^2_{\!\text{red}}$, so that we designate as $\text{min}\,\chi^2_{\!\text{red}}$ our measure of deviation of Eq.~(\ref{eq:apparent_V_1D}) from the simulations.

As $\sigma$ increases, the approximation fails above a value $\sigma_{\!\text{max}}$ that is determined by $\text{min}\,\chi^2_\text{red}=10$, as indicated in Fig.~\ref{fig6}c \cite{Bevington:2003tc}. The results for $\sigma_{\!\text{max}}$ are compared favorably to the solution of Eq.~(\ref{eq:maximum}a), shown by the black line in Fig.~\ref{fig6}c for various values of $k$. The same procedure is applied to evaluate a maximum static error $\varepsilon_\text{max}$ for each $k=1\dots7$ (Fig.~\ref{fig6}e), and compare it with the result of Eq.~(\ref{eq:maximum}b) shown by the black line (Fig.~\ref{fig6}f). We further verified that the latter results do not depend on $\sigma<\sigma_\text{max}$.

We have thus confirmed that Eqs.~(\ref{eq:condition}) provide effective estimates for the range of validity of Eq.~(\ref{eq:apparent_V}).

%%%%%%%%%%%%%%%%%%%%%%%%%%%%%%%%%%%%%%%%%%%%%%%%%%%%%%%%%%%%%%%%%%%%%%%%
%%%%%%%%%%%%%%%%%%%%%%%%%%%%%%%%%%%%%%%%%%%%%%%%%%%%%%%%%%%%%%%%%%%%%%%%


\begin{thebibliography}{57}%
\makeatletter
\providecommand \@ifxundefined [1]{%
 \@ifx{#1\undefined}
}%
\providecommand \@ifnum [1]{%
 \ifnum #1\expandafter \@firstoftwo
 \else \expandafter \@secondoftwo
 \fi
}%
\providecommand \@ifx [1]{%
 \ifx #1\expandafter \@firstoftwo
 \else \expandafter \@secondoftwo
 \fi
}%
\providecommand \natexlab [1]{#1}%
\providecommand \enquote  [1]{``#1''}%
\providecommand \bibnamefont  [1]{#1}%
\providecommand \bibfnamefont [1]{#1}%
\providecommand \citenamefont [1]{#1}%
\providecommand \href@noop [0]{\@secondoftwo}%
\providecommand \href [0]{\begingroup \@sanitize@url \@href}%
\providecommand \@href[1]{\@@startlink{#1}\@@href}%
\providecommand \@@href[1]{\endgroup#1\@@endlink}%
\providecommand \@sanitize@url [0]{\catcode `\\12\catcode `\$12\catcode
  `\&12\catcode `\#12\catcode `\^12\catcode `\_12\catcode `\%12\relax}%
\providecommand \@@startlink[1]{}%
\providecommand \@@endlink[0]{}%
\providecommand \url  [0]{\begingroup\@sanitize@url \@url }%
\providecommand \@url [1]{\endgroup\@href {#1}{\urlprefix }}%
\providecommand \urlprefix  [0]{URL }%
\providecommand \Eprint [0]{\href }%
\providecommand \doibase [0]{http://dx.doi.org/}%
\providecommand \selectlanguage [0]{\@gobble}%
\providecommand \bibinfo  [0]{\@secondoftwo}%
\providecommand \bibfield  [0]{\@secondoftwo}%
\providecommand \translation [1]{[#1]}%
\providecommand \BibitemOpen [0]{}%
\providecommand \bibitemStop [0]{}%
\providecommand \bibitemNoStop [0]{.\EOS\space}%
\providecommand \EOS [0]{\spacefactor3000\relax}%
\providecommand \BibitemShut  [1]{\csname bibitem#1\endcsname}%
\let\auto@bib@innerbib\@empty
%</preamble>
\bibitem [{\citenamefont {Meijering}\ \emph {et~al.}(2012)\citenamefont
  {Meijering}, \citenamefont {Dzyubachyk},\ and\ \citenamefont
  {Smal}}]{Meijering:2012hb}%
  \BibitemOpen
  \bibfield  {author} {\bibinfo {author} {\bibfnamefont {E.}~\bibnamefont
  {Meijering}}, \bibinfo {author} {\bibfnamefont {O.}~\bibnamefont
  {Dzyubachyk}}, \ and\ \bibinfo {author} {\bibfnamefont {I.}~\bibnamefont
  {Smal}},\ }in\ \href@noop {} {\emph {\bibinfo {booktitle} {Imaging and
  Spectroscopic Analysis of Living Cells - Optical and Spectroscopic
  Techniques}}}\ (\bibinfo  {publisher} {Elsevier},\ \bibinfo {year} {2012})\
  pp.\ \bibinfo {pages} {183--200}\BibitemShut {NoStop}%
\bibitem [{\citenamefont {Chenouard}\ \emph {et~al.}(2014)\citenamefont
  {Chenouard}, \citenamefont {Smal}, \citenamefont {de~Chaumont}, \citenamefont
  {Ma{\v s}ka}, \citenamefont {Sbalzarini}, \citenamefont {Gong}, \citenamefont
  {Cardinale}, \citenamefont {Carthel}, \citenamefont {Coraluppi},
  \citenamefont {Winter}, \citenamefont {Cohen}, \citenamefont {Godinez},
  \citenamefont {Rohr}, \citenamefont {Kalaidzidis}, \citenamefont {Liang},
  \citenamefont {Duncan}, \citenamefont {Shen}, \citenamefont {Xu},
  \citenamefont {Magnusson}, \citenamefont {Jald{\'e}n}, \citenamefont {Blau},
  \citenamefont {Paul-Gilloteaux}, \citenamefont {Roudot}, \citenamefont
  {Kervrann}, \citenamefont {Waharte}, \citenamefont {Tinevez}, \citenamefont
  {Shorte}, \citenamefont {Willemse}, \citenamefont {Celler}, \citenamefont
  {van Wezel}, \citenamefont {Dan}, \citenamefont {Tsai}, \citenamefont
  {de~Sol{\'o}rzano}, \citenamefont {Olivo-Marin},\ and\ \citenamefont
  {Meijering}}]{Chenouard:2014kg}%
  \BibitemOpen
  \bibfield  {author} {\bibinfo {author} {\bibfnamefont {N.}~\bibnamefont
  {Chenouard}}, \bibinfo {author} {\bibfnamefont {I.}~\bibnamefont {Smal}},
  \bibinfo {author} {\bibfnamefont {F.}~\bibnamefont {de~Chaumont}}, \bibinfo
  {author} {\bibfnamefont {M.}~\bibnamefont {Ma{\v s}ka}}, \bibinfo {author}
  {\bibfnamefont {I.~F.}\ \bibnamefont {Sbalzarini}}, \bibinfo {author}
  {\bibfnamefont {Y.}~\bibnamefont {Gong}}, \bibinfo {author} {\bibfnamefont
  {J.}~\bibnamefont {Cardinale}}, \bibinfo {author} {\bibfnamefont
  {C.}~\bibnamefont {Carthel}}, \bibinfo {author} {\bibfnamefont
  {S.}~\bibnamefont {Coraluppi}}, \bibinfo {author} {\bibfnamefont
  {M.}~\bibnamefont {Winter}}, \bibinfo {author} {\bibfnamefont {A.~R.}\
  \bibnamefont {Cohen}}, \bibinfo {author} {\bibfnamefont {W.~J.}\ \bibnamefont
  {Godinez}}, \bibinfo {author} {\bibfnamefont {K.}~\bibnamefont {Rohr}},
  \bibinfo {author} {\bibfnamefont {Y.}~\bibnamefont {Kalaidzidis}}, \bibinfo
  {author} {\bibfnamefont {L.}~\bibnamefont {Liang}}, \bibinfo {author}
  {\bibfnamefont {J.}~\bibnamefont {Duncan}}, \bibinfo {author} {\bibfnamefont
  {H.}~\bibnamefont {Shen}}, \bibinfo {author} {\bibfnamefont {Y.}~\bibnamefont
  {Xu}}, \bibinfo {author} {\bibfnamefont {K.~E.~G.}\ \bibnamefont
  {Magnusson}}, \bibinfo {author} {\bibfnamefont {J.}~\bibnamefont
  {Jald{\'e}n}}, \bibinfo {author} {\bibfnamefont {H.~M.}\ \bibnamefont
  {Blau}}, \bibinfo {author} {\bibfnamefont {P.}~\bibnamefont
  {Paul-Gilloteaux}}, \bibinfo {author} {\bibfnamefont {P.}~\bibnamefont
  {Roudot}}, \bibinfo {author} {\bibfnamefont {C.}~\bibnamefont {Kervrann}},
  \bibinfo {author} {\bibfnamefont {F.}~\bibnamefont {Waharte}}, \bibinfo
  {author} {\bibfnamefont {J.-Y.}\ \bibnamefont {Tinevez}}, \bibinfo {author}
  {\bibfnamefont {S.~L.}\ \bibnamefont {Shorte}}, \bibinfo {author}
  {\bibfnamefont {J.}~\bibnamefont {Willemse}}, \bibinfo {author}
  {\bibfnamefont {K.}~\bibnamefont {Celler}}, \bibinfo {author} {\bibfnamefont
  {G.~P.}\ \bibnamefont {van Wezel}}, \bibinfo {author} {\bibfnamefont {H.-W.}\
  \bibnamefont {Dan}}, \bibinfo {author} {\bibfnamefont {Y.-S.}\ \bibnamefont
  {Tsai}}, \bibinfo {author} {\bibfnamefont {C.~O.}\ \bibnamefont
  {de~Sol{\'o}rzano}}, \bibinfo {author} {\bibfnamefont {J.-C.}\ \bibnamefont
  {Olivo-Marin}}, \ and\ \bibinfo {author} {\bibfnamefont {E.}~\bibnamefont
  {Meijering}},\ }\href@noop {} {\bibfield  {journal} {\bibinfo  {journal}
  {Nat. Methods}\ }\textbf {\bibinfo {volume} {11}},\ \bibinfo {pages} {281}
  (\bibinfo {year} {2014})}\BibitemShut {NoStop}%
\bibitem [{\citenamefont {Manzo}\ and\ \citenamefont
  {Garcia-Parajo}(2015)}]{Manzo:2015dc}%
  \BibitemOpen
  \bibfield  {author} {\bibinfo {author} {\bibfnamefont {C.}~\bibnamefont
  {Manzo}}\ and\ \bibinfo {author} {\bibfnamefont {M.~F.}\ \bibnamefont
  {Garcia-Parajo}},\ }\href@noop {} {\bibfield  {journal} {\bibinfo  {journal}
  {Rep. Prog. Phys.}\ }\textbf {\bibinfo {volume} {78}},\ \bibinfo {pages}
  {124601} (\bibinfo {year} {2015})}\BibitemShut {NoStop}%
\bibitem [{\citenamefont {Courty}\ \emph {et~al.}(2006)\citenamefont {Courty},
  \citenamefont {Luccardini}, \citenamefont {Bellaiche}, \citenamefont
  {Cappello},\ and\ \citenamefont {Dahan}}]{Courty:2006km}%
  \BibitemOpen
  \bibfield  {author} {\bibinfo {author} {\bibfnamefont {S.}~\bibnamefont
  {Courty}}, \bibinfo {author} {\bibfnamefont {C.}~\bibnamefont {Luccardini}},
  \bibinfo {author} {\bibfnamefont {Y.}~\bibnamefont {Bellaiche}}, \bibinfo
  {author} {\bibfnamefont {G.}~\bibnamefont {Cappello}}, \ and\ \bibinfo
  {author} {\bibfnamefont {M.}~\bibnamefont {Dahan}},\ }\href@noop {}
  {\bibfield  {journal} {\bibinfo  {journal} {Nano Lett.}\ }\textbf {\bibinfo
  {volume} {6}},\ \bibinfo {pages} {1491} (\bibinfo {year} {2006})}\BibitemShut
  {NoStop}%
\bibitem [{\citenamefont {El~Beheiry}\ \emph {et~al.}(2015)\citenamefont
  {El~Beheiry}, \citenamefont {Dahan},\ and\ \citenamefont
  {Masson}}]{ElBeheiry:2015bv}%
  \BibitemOpen
  \bibfield  {author} {\bibinfo {author} {\bibfnamefont {M.}~\bibnamefont
  {El~Beheiry}}, \bibinfo {author} {\bibfnamefont {M.}~\bibnamefont {Dahan}}, \
  and\ \bibinfo {author} {\bibfnamefont {J.-B.}\ \bibnamefont {Masson}},\
  }\href@noop {} {\bibfield  {journal} {\bibinfo  {journal} {Nat. Neurosci.}\
  }\textbf {\bibinfo {volume} {12}},\ \bibinfo {pages} {594} (\bibinfo {year}
  {2015})}\BibitemShut {NoStop}%
\bibitem [{\citenamefont {Simson}\ \emph {et~al.}(1995)\citenamefont {Simson},
  \citenamefont {Sheets},\ and\ \citenamefont {Jacobson}}]{Simson:1995eh}%
  \BibitemOpen
  \bibfield  {author} {\bibinfo {author} {\bibfnamefont {R.}~\bibnamefont
  {Simson}}, \bibinfo {author} {\bibfnamefont {E.~D.}\ \bibnamefont {Sheets}},
  \ and\ \bibinfo {author} {\bibfnamefont {K.}~\bibnamefont {Jacobson}},\
  }\href@noop {} {\bibfield  {journal} {\bibinfo  {journal} {Biophys. J.}\
  }\textbf {\bibinfo {volume} {69}},\ \bibinfo {pages} {989} (\bibinfo {year}
  {1995})}\BibitemShut {NoStop}%
\bibitem [{\citenamefont {Mashanov}\ and\ \citenamefont
  {Molloy}(2007)}]{Mashanov:2007je}%
  \BibitemOpen
  \bibfield  {author} {\bibinfo {author} {\bibfnamefont {G.~I.}\ \bibnamefont
  {Mashanov}}\ and\ \bibinfo {author} {\bibfnamefont {J.~E.}\ \bibnamefont
  {Molloy}},\ }\href@noop {} {\bibfield  {journal} {\bibinfo  {journal}
  {Biophys. J.}\ }\textbf {\bibinfo {volume} {92}},\ \bibinfo {pages} {2199}
  (\bibinfo {year} {2007})}\BibitemShut {NoStop}%
\bibitem [{\citenamefont {Brandenburg}\ and\ \citenamefont
  {Zhuang}(2007)}]{Brandenburg:2007dt}%
  \BibitemOpen
  \bibfield  {author} {\bibinfo {author} {\bibfnamefont {B.}~\bibnamefont
  {Brandenburg}}\ and\ \bibinfo {author} {\bibfnamefont {X.}~\bibnamefont
  {Zhuang}},\ }\href@noop {} {\bibfield  {journal} {\bibinfo  {journal} {Nat
  Rev Micro}\ }\textbf {\bibinfo {volume} {5}},\ \bibinfo {pages} {197}
  (\bibinfo {year} {2007})}\BibitemShut {NoStop}%
\bibitem [{\citenamefont {Godinez}\ \emph {et~al.}(2009)\citenamefont
  {Godinez}, \citenamefont {Lampe}, \citenamefont {W{\"o}rz}, \citenamefont
  {M{\"u}ller}, \citenamefont {Eils},\ and\ \citenamefont
  {Rohr}}]{Godinez:2009cw}%
  \BibitemOpen
  \bibfield  {author} {\bibinfo {author} {\bibfnamefont {W.~J.}\ \bibnamefont
  {Godinez}}, \bibinfo {author} {\bibfnamefont {M.}~\bibnamefont {Lampe}},
  \bibinfo {author} {\bibfnamefont {S.}~\bibnamefont {W{\"o}rz}}, \bibinfo
  {author} {\bibfnamefont {B.}~\bibnamefont {M{\"u}ller}}, \bibinfo {author}
  {\bibfnamefont {R.}~\bibnamefont {Eils}}, \ and\ \bibinfo {author}
  {\bibfnamefont {K.}~\bibnamefont {Rohr}},\ }\href@noop {} {\bibfield
  {journal} {\bibinfo  {journal} {Medical Image Analysis}\ }\textbf {\bibinfo
  {volume} {13}},\ \bibinfo {pages} {325} (\bibinfo {year} {2009})}\BibitemShut
  {NoStop}%
\bibitem [{\citenamefont {Yasuda}\ \emph {et~al.}(1996)\citenamefont {Yasuda},
  \citenamefont {Miyata},\ and\ \citenamefont {Kinosita}}]{Yasuda:1996bt}%
  \BibitemOpen
  \bibfield  {author} {\bibinfo {author} {\bibfnamefont {R.}~\bibnamefont
  {Yasuda}}, \bibinfo {author} {\bibfnamefont {H.}~\bibnamefont {Miyata}}, \
  and\ \bibinfo {author} {\bibfnamefont {K.~J.}\ \bibnamefont {Kinosita}},\
  }\href@noop {} {\bibfield  {journal} {\bibinfo  {journal} {J. Mol. Biol.}\
  }\textbf {\bibinfo {volume} {263}},\ \bibinfo {pages} {227} (\bibinfo {year}
  {1996})}\BibitemShut {NoStop}%
\bibitem [{\citenamefont {Le~Goff}\ \emph {et~al.}(2002)\citenamefont
  {Le~Goff}, \citenamefont {Hallatschek}, \citenamefont {Frey},\ and\
  \citenamefont {Amblard}}]{LeGoff:2002bt}%
  \BibitemOpen
  \bibfield  {author} {\bibinfo {author} {\bibfnamefont {L.}~\bibnamefont
  {Le~Goff}}, \bibinfo {author} {\bibfnamefont {O.}~\bibnamefont
  {Hallatschek}}, \bibinfo {author} {\bibfnamefont {E.}~\bibnamefont {Frey}}, \
  and\ \bibinfo {author} {\bibfnamefont {F.}~\bibnamefont {Amblard}},\
  }\href@noop {} {\bibfield  {journal} {\bibinfo  {journal} {Phys. Rev. Lett.}\
  }\textbf {\bibinfo {volume} {89}},\ \bibinfo {pages} {258101} (\bibinfo
  {year} {2002})}\BibitemShut {NoStop}%
\bibitem [{\citenamefont {Jin}\ \emph {et~al.}(2007)\citenamefont {Jin},
  \citenamefont {Haggie},\ and\ \citenamefont {Verkman}}]{Jin:2007hq}%
  \BibitemOpen
  \bibfield  {author} {\bibinfo {author} {\bibfnamefont {S.}~\bibnamefont
  {Jin}}, \bibinfo {author} {\bibfnamefont {P.~M.}\ \bibnamefont {Haggie}}, \
  and\ \bibinfo {author} {\bibfnamefont {A.~S.}\ \bibnamefont {Verkman}},\
  }\href@noop {} {\bibfield  {journal} {\bibinfo  {journal} {Biophys. J.}\
  }\textbf {\bibinfo {volume} {93}},\ \bibinfo {pages} {1079} (\bibinfo {year}
  {2007})}\BibitemShut {NoStop}%
\bibitem [{\citenamefont {N{\"o}ding}\ and\ \citenamefont
  {K{\"o}ster}(2012)}]{Noding:2012km}%
  \BibitemOpen
  \bibfield  {author} {\bibinfo {author} {\bibfnamefont {B.}~\bibnamefont
  {N{\"o}ding}}\ and\ \bibinfo {author} {\bibfnamefont {S.}~\bibnamefont
  {K{\"o}ster}},\ }\href@noop {} {\bibfield  {journal} {\bibinfo  {journal}
  {Phys. Rev. Lett.}\ }\textbf {\bibinfo {volume} {108}},\ \bibinfo {pages}
  {088101} (\bibinfo {year} {2012})}\BibitemShut {NoStop}%
\bibitem [{\citenamefont {Dorfman}\ \emph {et~al.}(2013)\citenamefont
  {Dorfman}, \citenamefont {King}, \citenamefont {Olson}, \citenamefont
  {Thomas},\ and\ \citenamefont {Tree}}]{Dorfman:2013di}%
  \BibitemOpen
  \bibfield  {author} {\bibinfo {author} {\bibfnamefont {K.~D.}\ \bibnamefont
  {Dorfman}}, \bibinfo {author} {\bibfnamefont {S.~B.}\ \bibnamefont {King}},
  \bibinfo {author} {\bibfnamefont {D.~W.}\ \bibnamefont {Olson}}, \bibinfo
  {author} {\bibfnamefont {J.~D.~P.}\ \bibnamefont {Thomas}}, \ and\ \bibinfo
  {author} {\bibfnamefont {D.~R.}\ \bibnamefont {Tree}},\ }\href@noop {}
  {\bibfield  {journal} {\bibinfo  {journal} {Chem. Rev.}\ }\textbf {\bibinfo
  {volume} {113}},\ \bibinfo {pages} {2584} (\bibinfo {year}
  {2013})}\BibitemShut {NoStop}%
\bibitem [{\citenamefont {Engel}\ \emph {et~al.}(2014)\citenamefont {Engel},
  \citenamefont {Ritchie}, \citenamefont {Foster}, \citenamefont {Beach},\ and\
  \citenamefont {Woodside}}]{Engel:2014kq}%
  \BibitemOpen
  \bibfield  {author} {\bibinfo {author} {\bibfnamefont {M.~C.}\ \bibnamefont
  {Engel}}, \bibinfo {author} {\bibfnamefont {D.~B.}\ \bibnamefont {Ritchie}},
  \bibinfo {author} {\bibfnamefont {D.~A.~N.}\ \bibnamefont {Foster}}, \bibinfo
  {author} {\bibfnamefont {K.~S.~D.}\ \bibnamefont {Beach}}, \ and\ \bibinfo
  {author} {\bibfnamefont {M.~T.}\ \bibnamefont {Woodside}},\ }\href@noop {}
  {\bibfield  {journal} {\bibinfo  {journal} {Phys. Rev. Lett.}\ }\textbf
  {\bibinfo {volume} {113}},\ \bibinfo {pages} {238104} (\bibinfo {year}
  {2014})}\BibitemShut {NoStop}%
\bibitem [{\citenamefont {Hoze}\ \emph {et~al.}(2012)\citenamefont {Hoze},
  \citenamefont {Nair}, \citenamefont {Hosy}, \citenamefont {Sieben},
  \citenamefont {Manley}, \citenamefont {Herrmann}, \citenamefont {Sibarita},
  \citenamefont {Choquet},\ and\ \citenamefont {Holcman}}]{Hoze:2012jd}%
  \BibitemOpen
  \bibfield  {author} {\bibinfo {author} {\bibfnamefont {N.}~\bibnamefont
  {Hoze}}, \bibinfo {author} {\bibfnamefont {D.}~\bibnamefont {Nair}}, \bibinfo
  {author} {\bibfnamefont {E.}~\bibnamefont {Hosy}}, \bibinfo {author}
  {\bibfnamefont {C.}~\bibnamefont {Sieben}}, \bibinfo {author} {\bibfnamefont
  {S.}~\bibnamefont {Manley}}, \bibinfo {author} {\bibfnamefont
  {A.}~\bibnamefont {Herrmann}}, \bibinfo {author} {\bibfnamefont {J.~B.}\
  \bibnamefont {Sibarita}}, \bibinfo {author} {\bibfnamefont {D.}~\bibnamefont
  {Choquet}}, \ and\ \bibinfo {author} {\bibfnamefont {D.}~\bibnamefont
  {Holcman}},\ }\href@noop {} {\bibfield  {journal} {\bibinfo  {journal} {Proc
  Natl Acad Sci USA}\ }\textbf {\bibinfo {volume} {109}},\ \bibinfo {pages}
  {17052} (\bibinfo {year} {2012})}\BibitemShut {NoStop}%
\bibitem [{\citenamefont {Masson}\ \emph {et~al.}(2014)\citenamefont {Masson},
  \citenamefont {Dionne}, \citenamefont {Salvatico}, \citenamefont {Renner},
  \citenamefont {Specht}, \citenamefont {Triller},\ and\ \citenamefont
  {Dahan}}]{Masson:2014hr}%
  \BibitemOpen
  \bibfield  {author} {\bibinfo {author} {\bibfnamefont {J.-B.}\ \bibnamefont
  {Masson}}, \bibinfo {author} {\bibfnamefont {P.}~\bibnamefont {Dionne}},
  \bibinfo {author} {\bibfnamefont {C.}~\bibnamefont {Salvatico}}, \bibinfo
  {author} {\bibfnamefont {M.}~\bibnamefont {Renner}}, \bibinfo {author}
  {\bibfnamefont {C.~G.}\ \bibnamefont {Specht}}, \bibinfo {author}
  {\bibfnamefont {A.}~\bibnamefont {Triller}}, \ and\ \bibinfo {author}
  {\bibfnamefont {M.}~\bibnamefont {Dahan}},\ }\href@noop {} {\bibfield
  {journal} {\bibinfo  {journal} {Biophys. J.}\ }\textbf {\bibinfo {volume}
  {106}},\ \bibinfo {pages} {74} (\bibinfo {year} {2014})}\BibitemShut
  {NoStop}%
\bibitem [{\citenamefont {Qian}\ \emph {et~al.}(1991)\citenamefont {Qian},
  \citenamefont {Sheetz},\ and\ \citenamefont {Elson}}]{Qian:1991vy}%
  \BibitemOpen
  \bibfield  {author} {\bibinfo {author} {\bibfnamefont {H.}~\bibnamefont
  {Qian}}, \bibinfo {author} {\bibfnamefont {M.~P.}\ \bibnamefont {Sheetz}}, \
  and\ \bibinfo {author} {\bibfnamefont {E.~L.}\ \bibnamefont {Elson}},\
  }\href@noop {} {\bibfield  {journal} {\bibinfo  {journal} {Biophys. J.}\
  }\textbf {\bibinfo {volume} {60}},\ \bibinfo {pages} {910} (\bibinfo {year}
  {1991})}\BibitemShut {NoStop}%
\bibitem [{\citenamefont {Saxton}(1997)}]{Saxton:1997uka}%
  \BibitemOpen
  \bibfield  {author} {\bibinfo {author} {\bibfnamefont {M.~J.}\ \bibnamefont
  {Saxton}},\ }\href@noop {} {\bibfield  {journal} {\bibinfo  {journal}
  {Biophys. J.}\ }\textbf {\bibinfo {volume} {72}},\ \bibinfo {pages} {1744}
  (\bibinfo {year} {1997})}\BibitemShut {NoStop}%
\bibitem [{\citenamefont {Fatin-Rouge}\ \emph {et~al.}(2004)\citenamefont
  {Fatin-Rouge}, \citenamefont {Starchev},\ and\ \citenamefont
  {Buffle}}]{FatinRouge:2004dt}%
  \BibitemOpen
  \bibfield  {author} {\bibinfo {author} {\bibfnamefont {N.}~\bibnamefont
  {Fatin-Rouge}}, \bibinfo {author} {\bibfnamefont {K.}~\bibnamefont
  {Starchev}}, \ and\ \bibinfo {author} {\bibfnamefont {J.}~\bibnamefont
  {Buffle}},\ }\href@noop {} {\bibfield  {journal} {\bibinfo  {journal}
  {Biophys. J.}\ }\textbf {\bibinfo {volume} {86}},\ \bibinfo {pages} {2710}
  (\bibinfo {year} {2004})}\BibitemShut {NoStop}%
\bibitem [{\citenamefont {Waigh}(2016)}]{Waigh:2016kh}%
  \BibitemOpen
  \bibfield  {author} {\bibinfo {author} {\bibfnamefont {T.~A.}\ \bibnamefont
  {Waigh}},\ }\href@noop {} {\bibfield  {journal} {\bibinfo  {journal} {Rep.
  Prog. Phys.}\ }\textbf {\bibinfo {volume} {79}},\ \bibinfo {pages} {1}
  (\bibinfo {year} {2016})}\BibitemShut {NoStop}%
\bibitem [{\citenamefont {Crocker}\ and\ \citenamefont
  {Grier}(1998)}]{Crocker:1998gp}%
  \BibitemOpen
  \bibfield  {author} {\bibinfo {author} {\bibfnamefont {J.~C.}\ \bibnamefont
  {Crocker}}\ and\ \bibinfo {author} {\bibfnamefont {D.~G.}\ \bibnamefont
  {Grier}},\ }\href@noop {} {\bibfield  {journal} {\bibinfo  {journal} {MRS
  Bull.}\ }\textbf {\bibinfo {volume} {23}},\ \bibinfo {pages} {24} (\bibinfo
  {year} {1998})}\BibitemShut {NoStop}%
\bibitem [{\citenamefont {Crocker}\ and\ \citenamefont
  {Grier}(1994)}]{Crocker:1994ge}%
  \BibitemOpen
  \bibfield  {author} {\bibinfo {author} {\bibfnamefont {J.~C.}\ \bibnamefont
  {Crocker}}\ and\ \bibinfo {author} {\bibfnamefont {D.~G.}\ \bibnamefont
  {Grier}},\ }\href@noop {} {\bibfield  {journal} {\bibinfo  {journal} {Phys.
  Rev. Lett.}\ }\textbf {\bibinfo {volume} {73}},\ \bibinfo {pages} {352}
  (\bibinfo {year} {1994})}\BibitemShut {NoStop}%
\bibitem [{\citenamefont {Crocker}\ \emph {et~al.}(1999)\citenamefont
  {Crocker}, \citenamefont {Matteo}, \citenamefont {Dinsmore},\ and\
  \citenamefont {Yodh}}]{Crocker:1999fq}%
  \BibitemOpen
  \bibfield  {author} {\bibinfo {author} {\bibfnamefont {J.~C.}\ \bibnamefont
  {Crocker}}, \bibinfo {author} {\bibfnamefont {J.~A.}\ \bibnamefont {Matteo}},
  \bibinfo {author} {\bibfnamefont {A.~D.}\ \bibnamefont {Dinsmore}}, \ and\
  \bibinfo {author} {\bibfnamefont {A.~G.}\ \bibnamefont {Yodh}},\ }\href@noop
  {} {\bibfield  {journal} {\bibinfo  {journal} {Phys. Rev. Lett.}\ }\textbf
  {\bibinfo {volume} {82}},\ \bibinfo {pages} {4352} (\bibinfo {year}
  {1999})}\BibitemShut {NoStop}%
\bibitem [{\citenamefont {Lin}\ \emph {et~al.}(2001)\citenamefont {Lin},
  \citenamefont {Crocker}, \citenamefont {Zeri},\ and\ \citenamefont
  {Yodh}}]{Lin:2001gt}%
  \BibitemOpen
  \bibfield  {author} {\bibinfo {author} {\bibfnamefont {K.-H.}\ \bibnamefont
  {Lin}}, \bibinfo {author} {\bibfnamefont {J.~C.}\ \bibnamefont {Crocker}},
  \bibinfo {author} {\bibfnamefont {A.~C.}\ \bibnamefont {Zeri}}, \ and\
  \bibinfo {author} {\bibfnamefont {A.}~\bibnamefont {Yodh}},\ }\href@noop {}
  {\bibfield  {journal} {\bibinfo  {journal} {Phys. Rev. Lett.}\ }\textbf
  {\bibinfo {volume} {87}},\ \bibinfo {pages} {088301} (\bibinfo {year}
  {2001})}\BibitemShut {NoStop}%
\bibitem [{\citenamefont {Mojarad}\ and\ \citenamefont
  {Krishnan}(2012)}]{Mojarad:2012js}%
  \BibitemOpen
  \bibfield  {author} {\bibinfo {author} {\bibfnamefont {N.}~\bibnamefont
  {Mojarad}}\ and\ \bibinfo {author} {\bibfnamefont {M.}~\bibnamefont
  {Krishnan}},\ }\href@noop {} {\bibfield  {journal} {\bibinfo  {journal} {Nat.
  Nanotechnol.}\ }\textbf {\bibinfo {volume} {7}},\ \bibinfo {pages} {448}
  (\bibinfo {year} {2012})}\BibitemShut {NoStop}%
\bibitem [{\citenamefont {Krishnan}\ \emph {et~al.}(2010)\citenamefont
  {Krishnan}, \citenamefont {Mojarad}, \citenamefont {Kukura},\ and\
  \citenamefont {Sandoghdar}}]{Krishnan:2010fp}%
  \BibitemOpen
  \bibfield  {author} {\bibinfo {author} {\bibfnamefont {M.}~\bibnamefont
  {Krishnan}}, \bibinfo {author} {\bibfnamefont {N.}~\bibnamefont {Mojarad}},
  \bibinfo {author} {\bibfnamefont {P.}~\bibnamefont {Kukura}}, \ and\ \bibinfo
  {author} {\bibfnamefont {V.}~\bibnamefont {Sandoghdar}},\ }\href@noop {}
  {\bibfield  {journal} {\bibinfo  {journal} {Nature}\ }\textbf {\bibinfo
  {volume} {467}},\ \bibinfo {pages} {692} (\bibinfo {year}
  {2010})}\BibitemShut {NoStop}%
\bibitem [{\citenamefont {Mojarad}\ \emph {et~al.}(2013)\citenamefont
  {Mojarad}, \citenamefont {Sandoghdar},\ and\ \citenamefont
  {Krishnan}}]{Mojarad:2013hu}%
  \BibitemOpen
  \bibfield  {author} {\bibinfo {author} {\bibfnamefont {N.}~\bibnamefont
  {Mojarad}}, \bibinfo {author} {\bibfnamefont {V.}~\bibnamefont {Sandoghdar}},
  \ and\ \bibinfo {author} {\bibfnamefont {M.}~\bibnamefont {Krishnan}},\
  }\href@noop {} {\bibfield  {journal} {\bibinfo  {journal} {Opt. Express}\
  }\textbf {\bibinfo {volume} {21}},\ \bibinfo {pages} {9377} (\bibinfo {year}
  {2013})}\BibitemShut {NoStop}%
\bibitem [{\citenamefont {Pagliara}\ \emph {et~al.}(2013)\citenamefont
  {Pagliara}, \citenamefont {Schwall},\ and\ \citenamefont
  {Keyser}}]{Pagliara:2013gi}%
  \BibitemOpen
  \bibfield  {author} {\bibinfo {author} {\bibfnamefont {S.}~\bibnamefont
  {Pagliara}}, \bibinfo {author} {\bibfnamefont {C.}~\bibnamefont {Schwall}}, \
  and\ \bibinfo {author} {\bibfnamefont {U.~F.}\ \bibnamefont {Keyser}},\
  }\href@noop {} {\bibfield  {journal} {\bibinfo  {journal} {Adv. Mater.}\
  }\textbf {\bibinfo {volume} {25}},\ \bibinfo {pages} {844} (\bibinfo {year}
  {2013})}\BibitemShut {NoStop}%
\bibitem [{\citenamefont {Lee}\ \emph {et~al.}(2017)\citenamefont {Lee},
  \citenamefont {Tsekouras}, \citenamefont {Calderon}, \citenamefont
  {Bustamante},\ and\ \citenamefont {Press{\'e}}}]{Lee:2017fb}%
  \BibitemOpen
  \bibfield  {author} {\bibinfo {author} {\bibfnamefont {A.}~\bibnamefont
  {Lee}}, \bibinfo {author} {\bibfnamefont {K.}~\bibnamefont {Tsekouras}},
  \bibinfo {author} {\bibfnamefont {C.}~\bibnamefont {Calderon}}, \bibinfo
  {author} {\bibfnamefont {C.}~\bibnamefont {Bustamante}}, \ and\ \bibinfo
  {author} {\bibfnamefont {S.}~\bibnamefont {Press{\'e}}},\ }\href@noop {}
  {\bibfield  {journal} {\bibinfo  {journal} {Chem. Rev.}\ }\textbf {\bibinfo
  {volume} {117}},\ \bibinfo {pages} {7276} (\bibinfo {year}
  {2017})}\BibitemShut {NoStop}%
\bibitem [{\citenamefont {Saxton}\ and\ \citenamefont
  {Jacobson}(1997)}]{Saxton:1997uk}%
  \BibitemOpen
  \bibfield  {author} {\bibinfo {author} {\bibfnamefont {M.~J.}\ \bibnamefont
  {Saxton}}\ and\ \bibinfo {author} {\bibfnamefont {K.}~\bibnamefont
  {Jacobson}},\ }\href@noop {} {\bibfield  {journal} {\bibinfo  {journal}
  {Annu. Rev. Biophys. Biomol. Struct.}\ }\textbf {\bibinfo {volume} {26}},\
  \bibinfo {pages} {373} (\bibinfo {year} {1997})}\BibitemShut {NoStop}%
\bibitem [{\citenamefont {Oddershede}\ \emph {et~al.}(2002)\citenamefont
  {Oddershede}, \citenamefont {Dreyer}, \citenamefont {Grego}, \citenamefont
  {Brown},\ and\ \citenamefont {Berg-S{\o}rensen}}]{Oddershede:2002gu}%
  \BibitemOpen
  \bibfield  {author} {\bibinfo {author} {\bibfnamefont {L.}~\bibnamefont
  {Oddershede}}, \bibinfo {author} {\bibfnamefont {J.~K.}\ \bibnamefont
  {Dreyer}}, \bibinfo {author} {\bibfnamefont {S.}~\bibnamefont {Grego}},
  \bibinfo {author} {\bibfnamefont {S.}~\bibnamefont {Brown}}, \ and\ \bibinfo
  {author} {\bibfnamefont {K.}~\bibnamefont {Berg-S{\o}rensen}},\ }\href@noop
  {} {\bibfield  {journal} {\bibinfo  {journal} {Biophys. J.}\ }\textbf
  {\bibinfo {volume} {83}},\ \bibinfo {pages} {3152} (\bibinfo {year}
  {2002})}\BibitemShut {NoStop}%
\bibitem [{\citenamefont {Jenkins}\ \emph {et~al.}(2015)\citenamefont
  {Jenkins}, \citenamefont {Crocker},\ and\ \citenamefont
  {Sinno}}]{Jenkins:2015ida}%
  \BibitemOpen
  \bibfield  {author} {\bibinfo {author} {\bibfnamefont {I.~C.}\ \bibnamefont
  {Jenkins}}, \bibinfo {author} {\bibfnamefont {J.~C.}\ \bibnamefont
  {Crocker}}, \ and\ \bibinfo {author} {\bibfnamefont {T.}~\bibnamefont
  {Sinno}},\ }\href@noop {} {\bibfield  {journal} {\bibinfo  {journal} {Soft
  Matter}\ }\textbf {\bibinfo {volume} {11}},\ \bibinfo {pages} {6948}
  (\bibinfo {year} {2015})}\BibitemShut {NoStop}%
\bibitem [{\citenamefont {Cheezum}\ \emph {et~al.}(2001)\citenamefont
  {Cheezum}, \citenamefont {Walker},\ and\ \citenamefont
  {Guilford}}]{Cheezum:2001ga}%
  \BibitemOpen
  \bibfield  {author} {\bibinfo {author} {\bibfnamefont {M.~K.}\ \bibnamefont
  {Cheezum}}, \bibinfo {author} {\bibfnamefont {W.~F.}\ \bibnamefont {Walker}},
  \ and\ \bibinfo {author} {\bibfnamefont {W.~H.}\ \bibnamefont {Guilford}},\
  }\href@noop {} {\bibfield  {journal} {\bibinfo  {journal} {Biophys. J.}\
  }\textbf {\bibinfo {volume} {81}},\ \bibinfo {pages} {2378} (\bibinfo {year}
  {2001})}\BibitemShut {NoStop}%
\bibitem [{\citenamefont {Smal}\ \emph {et~al.}(2008)\citenamefont {Smal},
  \citenamefont {Draegestein}, \citenamefont {Galjart}, \citenamefont
  {Niessen},\ and\ \citenamefont {Meijering}}]{Smal:2008dy}%
  \BibitemOpen
  \bibfield  {author} {\bibinfo {author} {\bibfnamefont {I.}~\bibnamefont
  {Smal}}, \bibinfo {author} {\bibfnamefont {K.}~\bibnamefont {Draegestein}},
  \bibinfo {author} {\bibfnamefont {N.}~\bibnamefont {Galjart}}, \bibinfo
  {author} {\bibfnamefont {W.}~\bibnamefont {Niessen}}, \ and\ \bibinfo
  {author} {\bibfnamefont {E.}~\bibnamefont {Meijering}},\ }\href@noop {}
  {\bibfield  {journal} {\bibinfo  {journal} {IEEE Trans. Med. Imaging}\
  }\textbf {\bibinfo {volume} {27}},\ \bibinfo {pages} {789} (\bibinfo {year}
  {2008})}\BibitemShut {NoStop}%
\bibitem [{\citenamefont {Martin}\ \emph {et~al.}(2002)\citenamefont {Martin},
  \citenamefont {Forstner},\ and\ \citenamefont {Kas}}]{Martin:2002ch}%
  \BibitemOpen
  \bibfield  {author} {\bibinfo {author} {\bibfnamefont {D.~S.}\ \bibnamefont
  {Martin}}, \bibinfo {author} {\bibfnamefont {M.~B.}\ \bibnamefont
  {Forstner}}, \ and\ \bibinfo {author} {\bibfnamefont {J.~A.}\ \bibnamefont
  {Kas}},\ }\href@noop {} {\bibfield  {journal} {\bibinfo  {journal} {Biophys.
  J.}\ }\textbf {\bibinfo {volume} {83}},\ \bibinfo {pages} {2109} (\bibinfo
  {year} {2002})}\BibitemShut {NoStop}%
\bibitem [{\citenamefont {Savin}\ and\ \citenamefont
  {Doyle}(2005{\natexlab{a}})}]{Savin:2005ko}%
  \BibitemOpen
  \bibfield  {author} {\bibinfo {author} {\bibfnamefont {T.}~\bibnamefont
  {Savin}}\ and\ \bibinfo {author} {\bibfnamefont {P.~S.}\ \bibnamefont
  {Doyle}},\ }\href@noop {} {\bibfield  {journal} {\bibinfo  {journal} {Phys.
  Rev. E}\ }\textbf {\bibinfo {volume} {71}},\ \bibinfo {pages} {041106}
  (\bibinfo {year} {2005}{\natexlab{a}})}\BibitemShut {NoStop}%
\bibitem [{\citenamefont {Savin}\ and\ \citenamefont
  {Doyle}(2005{\natexlab{b}})}]{Savin:2005ch}%
  \BibitemOpen
  \bibfield  {author} {\bibinfo {author} {\bibfnamefont {T.}~\bibnamefont
  {Savin}}\ and\ \bibinfo {author} {\bibfnamefont {P.~S.}\ \bibnamefont
  {Doyle}},\ }\href@noop {} {\bibfield  {journal} {\bibinfo  {journal}
  {Biophys. J.}\ }\textbf {\bibinfo {volume} {88}},\ \bibinfo {pages} {623}
  (\bibinfo {year} {2005}{\natexlab{b}})}\BibitemShut {NoStop}%
\bibitem [{\citenamefont {Ritchie}\ \emph {et~al.}(2005)\citenamefont
  {Ritchie}, \citenamefont {Shan}, \citenamefont {Kondo}, \citenamefont
  {Iwasawa}, \citenamefont {Fujiwara},\ and\ \citenamefont
  {Kusumi}}]{Ritchie:2005fj}%
  \BibitemOpen
  \bibfield  {author} {\bibinfo {author} {\bibfnamefont {K.}~\bibnamefont
  {Ritchie}}, \bibinfo {author} {\bibfnamefont {X.-Y.}\ \bibnamefont {Shan}},
  \bibinfo {author} {\bibfnamefont {J.}~\bibnamefont {Kondo}}, \bibinfo
  {author} {\bibfnamefont {K.}~\bibnamefont {Iwasawa}}, \bibinfo {author}
  {\bibfnamefont {T.}~\bibnamefont {Fujiwara}}, \ and\ \bibinfo {author}
  {\bibfnamefont {A.}~\bibnamefont {Kusumi}},\ }\href@noop {} {\bibfield
  {journal} {\bibinfo  {journal} {Biophys. J.}\ }\textbf {\bibinfo {volume}
  {88}},\ \bibinfo {pages} {2266} (\bibinfo {year} {2005})}\BibitemShut
  {NoStop}%
\bibitem [{\citenamefont {Wong}\ and\ \citenamefont
  {Halvorsen}(2006)}]{Wong:2006da}%
  \BibitemOpen
  \bibfield  {author} {\bibinfo {author} {\bibfnamefont {W.~P.}\ \bibnamefont
  {Wong}}\ and\ \bibinfo {author} {\bibfnamefont {K.}~\bibnamefont
  {Halvorsen}},\ }\href@noop {} {\bibfield  {journal} {\bibinfo  {journal}
  {Opt. Express}\ }\textbf {\bibinfo {volume} {14}},\ \bibinfo {pages} {12517}
  (\bibinfo {year} {2006})}\BibitemShut {NoStop}%
\bibitem [{\citenamefont {van~der Horst}\ and\ \citenamefont
  {Forde}(2010)}]{vanderHorst:2010kt}%
  \BibitemOpen
  \bibfield  {author} {\bibinfo {author} {\bibfnamefont {A.}~\bibnamefont
  {van~der Horst}}\ and\ \bibinfo {author} {\bibfnamefont {N.~R.}\ \bibnamefont
  {Forde}},\ }\href@noop {} {\bibfield  {journal} {\bibinfo  {journal} {Opt.
  Express}\ }\textbf {\bibinfo {volume} {18}},\ \bibinfo {pages} {7670}
  (\bibinfo {year} {2010})}\BibitemShut {NoStop}%
\bibitem [{\citenamefont {Michalet}(2010)}]{Michalet:2010hd}%
  \BibitemOpen
  \bibfield  {author} {\bibinfo {author} {\bibfnamefont {X.}~\bibnamefont
  {Michalet}},\ }\href@noop {} {\bibfield  {journal} {\bibinfo  {journal}
  {Phys. Rev. E}\ }\textbf {\bibinfo {volume} {82}},\ \bibinfo {pages} {041914}
  (\bibinfo {year} {2010})}\BibitemShut {NoStop}%
\bibitem [{\citenamefont {Berglund}(2010)}]{Berglund:2010ff}%
  \BibitemOpen
  \bibfield  {author} {\bibinfo {author} {\bibfnamefont {A.~J.}\ \bibnamefont
  {Berglund}},\ }\href@noop {} {\bibfield  {journal} {\bibinfo  {journal}
  {Phys. Rev. E}\ }\textbf {\bibinfo {volume} {82}},\ \bibinfo {pages} {011917}
  (\bibinfo {year} {2010})}\BibitemShut {NoStop}%
\bibitem [{\citenamefont {Hoze}\ and\ \citenamefont
  {Holcman}(2015)}]{Hoze:2015jo}%
  \BibitemOpen
  \bibfield  {author} {\bibinfo {author} {\bibfnamefont {N.}~\bibnamefont
  {Hoze}}\ and\ \bibinfo {author} {\bibfnamefont {D.}~\bibnamefont {Holcman}},\
  }\href@noop {} {\bibfield  {journal} {\bibinfo  {journal} {Phys. Rev. E}\
  }\textbf {\bibinfo {volume} {92}},\ \bibinfo {pages} {052109} (\bibinfo
  {year} {2015})}\BibitemShut {NoStop}%
\bibitem [{\citenamefont {Calderon}(2016)}]{Calderon:2016ia}%
  \BibitemOpen
  \bibfield  {author} {\bibinfo {author} {\bibfnamefont {C.~P.}\ \bibnamefont
  {Calderon}},\ }\href@noop {} {\bibfield  {journal} {\bibinfo  {journal}
  {Phys. Rev. E}\ }\textbf {\bibinfo {volume} {93}},\ \bibinfo {pages} {053303}
  (\bibinfo {year} {2016})}\BibitemShut {NoStop}%
\bibitem [{\citenamefont {Burov}\ \emph {et~al.}(2017)\citenamefont {Burov},
  \citenamefont {Figliozzi}, \citenamefont {Lin}, \citenamefont {Rice},
  \citenamefont {Scherer},\ and\ \citenamefont {Dinner}}]{Burov:2017bt}%
  \BibitemOpen
  \bibfield  {author} {\bibinfo {author} {\bibfnamefont {S.}~\bibnamefont
  {Burov}}, \bibinfo {author} {\bibfnamefont {P.}~\bibnamefont {Figliozzi}},
  \bibinfo {author} {\bibfnamefont {B.}~\bibnamefont {Lin}}, \bibinfo {author}
  {\bibfnamefont {S.~A.}\ \bibnamefont {Rice}}, \bibinfo {author}
  {\bibfnamefont {N.~F.}\ \bibnamefont {Scherer}}, \ and\ \bibinfo {author}
  {\bibfnamefont {A.~R.}\ \bibnamefont {Dinner}},\ }\href@noop {} {\bibfield
  {journal} {\bibinfo  {journal} {Proc. Natl. Acad. Sci. U.S.A.}\ }\textbf
  {\bibinfo {volume} {114}},\ \bibinfo {pages} {221} (\bibinfo {year}
  {2017})}\BibitemShut {NoStop}%
\bibitem [{\citenamefont {Hoz{\'e}}\ and\ \citenamefont
  {Holcman}(2017)}]{Hoze:2017bc}%
  \BibitemOpen
  \bibfield  {author} {\bibinfo {author} {\bibfnamefont {N.}~\bibnamefont
  {Hoz{\'e}}}\ and\ \bibinfo {author} {\bibfnamefont {D.}~\bibnamefont
  {Holcman}},\ }\href@noop {} {\bibfield  {journal} {\bibinfo  {journal} {Annu.
  Rev. Stat. Appl.}\ }\textbf {\bibinfo {volume} {4}},\ \bibinfo {pages} {189}
  (\bibinfo {year} {2017})}\BibitemShut {NoStop}%
\bibitem [{\citenamefont {Savin}\ \emph {et~al.}(2008)\citenamefont {Savin},
  \citenamefont {Spicer},\ and\ \citenamefont {Doyle}}]{Savin:2008dv}%
  \BibitemOpen
  \bibfield  {author} {\bibinfo {author} {\bibfnamefont {T.}~\bibnamefont
  {Savin}}, \bibinfo {author} {\bibfnamefont {P.~T.}\ \bibnamefont {Spicer}}, \
  and\ \bibinfo {author} {\bibfnamefont {P.~S.}\ \bibnamefont {Doyle}},\
  }\href@noop {} {\bibfield  {journal} {\bibinfo  {journal} {Appl. Phys.
  Lett.}\ }\textbf {\bibinfo {volume} {93}},\ \bibinfo {pages} {024102}
  (\bibinfo {year} {2008})}\BibitemShut {NoStop}%
\bibitem [{\citenamefont {T{\"u}rkcan}\ \emph {et~al.}(2012)\citenamefont
  {T{\"u}rkcan}, \citenamefont {Alexandrou},\ and\ \citenamefont
  {Masson}}]{Turkcan:2012ev}%
  \BibitemOpen
  \bibfield  {author} {\bibinfo {author} {\bibfnamefont {S.}~\bibnamefont
  {T{\"u}rkcan}}, \bibinfo {author} {\bibfnamefont {A.}~\bibnamefont
  {Alexandrou}}, \ and\ \bibinfo {author} {\bibfnamefont {J.-B.}\ \bibnamefont
  {Masson}},\ }\href@noop {} {\bibfield  {journal} {\bibinfo  {journal}
  {Biophys. J.}\ }\textbf {\bibinfo {volume} {102}},\ \bibinfo {pages} {2288}
  (\bibinfo {year} {2012})}\BibitemShut {NoStop}%
\bibitem [{\citenamefont {El~Beheiry}\ \emph {et~al.}(2016)\citenamefont
  {El~Beheiry}, \citenamefont {T{\"u}rkcan}, \citenamefont {Richly},
  \citenamefont {Triller}, \citenamefont {Alexandrou}, \citenamefont {Dahan},\
  and\ \citenamefont {Masson}}]{ElBeheiry:2016jg}%
  \BibitemOpen
  \bibfield  {author} {\bibinfo {author} {\bibfnamefont {M.}~\bibnamefont
  {El~Beheiry}}, \bibinfo {author} {\bibfnamefont {S.}~\bibnamefont
  {T{\"u}rkcan}}, \bibinfo {author} {\bibfnamefont {M.~U.}\ \bibnamefont
  {Richly}}, \bibinfo {author} {\bibfnamefont {A.}~\bibnamefont {Triller}},
  \bibinfo {author} {\bibfnamefont {A.}~\bibnamefont {Alexandrou}}, \bibinfo
  {author} {\bibfnamefont {M.}~\bibnamefont {Dahan}}, \ and\ \bibinfo {author}
  {\bibfnamefont {J.-B.}\ \bibnamefont {Masson}},\ }\href@noop {} {\bibfield
  {journal} {\bibinfo  {journal} {Biophys. J.}\ }\textbf {\bibinfo {volume}
  {110}},\ \bibinfo {pages} {1209} (\bibinfo {year} {2016})}\BibitemShut
  {NoStop}%
\bibitem [{\citenamefont {Keller}\ \emph {et~al.}(2001)\citenamefont {Keller},
  \citenamefont {Schilling},\ and\ \citenamefont {Sackmann}}]{Keller:2001dw}%
  \BibitemOpen
  \bibfield  {author} {\bibinfo {author} {\bibfnamefont {M.}~\bibnamefont
  {Keller}}, \bibinfo {author} {\bibfnamefont {J.}~\bibnamefont {Schilling}}, \
  and\ \bibinfo {author} {\bibfnamefont {E.}~\bibnamefont {Sackmann}},\
  }\href@noop {} {\bibfield  {journal} {\bibinfo  {journal} {Rev. Sci.
  Instrum.}\ }\textbf {\bibinfo {volume} {72}},\ \bibinfo {pages} {3626}
  (\bibinfo {year} {2001})}\BibitemShut {NoStop}%
\bibitem [{\citenamefont {{\"O}ttinger}(1996)}]{Ottinger:1996gx}%
  \BibitemOpen
  \bibfield  {author} {\bibinfo {author} {\bibfnamefont {H.~C.}\ \bibnamefont
  {{\"O}ttinger}},\ }\href@noop {} {\emph {\bibinfo {title} {{Stochastic
  Processes in Polymeric Fluids: Tools and Examples for Developing Simulation
  Algorithms}}}}\ (\bibinfo  {publisher} {Springer},\ \bibinfo {address}
  {Berlin},\ \bibinfo {year} {1996})\BibitemShut {NoStop}%
\bibitem [{\citenamefont {Israelachvili}(2011)}]{Israelachvili:2011ug}%
  \BibitemOpen
  \bibfield  {author} {\bibinfo {author} {\bibfnamefont {J.~N.}\ \bibnamefont
  {Israelachvili}},\ }\href@noop {} {\emph {\bibinfo {title} {{Intermolecular
  and Surface Forces}}}},\ \bibinfo {edition} {3rd}\ ed.\ (\bibinfo
  {publisher} {Academic Press},\ \bibinfo {address} {San Diego},\ \bibinfo
  {year} {2011})\BibitemShut {NoStop}%
\bibitem [{\citenamefont {Lau}\ \emph {et~al.}(2002)\citenamefont {Lau},
  \citenamefont {Lin},\ and\ \citenamefont {Yodh}}]{Lau:2002ft}%
  \BibitemOpen
  \bibfield  {author} {\bibinfo {author} {\bibfnamefont {A.}~\bibnamefont
  {Lau}}, \bibinfo {author} {\bibfnamefont {K.-H.}\ \bibnamefont {Lin}}, \ and\
  \bibinfo {author} {\bibfnamefont {A.}~\bibnamefont {Yodh}},\ }\href@noop {}
  {\bibfield  {journal} {\bibinfo  {journal} {Phys. Rev. E}\ }\textbf {\bibinfo
  {volume} {66}},\ \bibinfo {pages} {020401} (\bibinfo {year}
  {2002})}\BibitemShut {NoStop}%
\bibitem [{\citenamefont {Valentine}\ \emph {et~al.}(2001)\citenamefont
  {Valentine}, \citenamefont {Kaplan}, \citenamefont {Thota}, \citenamefont
  {Crocker}, \citenamefont {Gisler}, \citenamefont {Prud'homme}, \citenamefont
  {Beck},\ and\ \citenamefont {Weitz}}]{Valentine:2001fx}%
  \BibitemOpen
  \bibfield  {author} {\bibinfo {author} {\bibfnamefont {M.~T.}\ \bibnamefont
  {Valentine}}, \bibinfo {author} {\bibfnamefont {P.~D.}\ \bibnamefont
  {Kaplan}}, \bibinfo {author} {\bibfnamefont {D.}~\bibnamefont {Thota}},
  \bibinfo {author} {\bibfnamefont {J.~C.}\ \bibnamefont {Crocker}}, \bibinfo
  {author} {\bibfnamefont {T.}~\bibnamefont {Gisler}}, \bibinfo {author}
  {\bibfnamefont {R.~K.}\ \bibnamefont {Prud'homme}}, \bibinfo {author}
  {\bibfnamefont {M.}~\bibnamefont {Beck}}, \ and\ \bibinfo {author}
  {\bibfnamefont {D.~A.}\ \bibnamefont {Weitz}},\ }\href@noop {} {\bibfield
  {journal} {\bibinfo  {journal} {Phys. Rev. E}\ }\textbf {\bibinfo {volume}
  {64}},\ \bibinfo {pages} {061506} (\bibinfo {year} {2001})}\BibitemShut
  {NoStop}%
\bibitem [{\citenamefont {Crocker}\ \emph {et~al.}(2000)\citenamefont
  {Crocker}, \citenamefont {Valentine}, \citenamefont {Weeks}, \citenamefont
  {Gisler}, \citenamefont {Kaplan}, \citenamefont {Yodh},\ and\ \citenamefont
  {Weitz}}]{Crocker:2000ct}%
  \BibitemOpen
  \bibfield  {author} {\bibinfo {author} {\bibfnamefont {J.~C.}\ \bibnamefont
  {Crocker}}, \bibinfo {author} {\bibfnamefont {M.~T.}\ \bibnamefont
  {Valentine}}, \bibinfo {author} {\bibfnamefont {E.~R.}\ \bibnamefont
  {Weeks}}, \bibinfo {author} {\bibfnamefont {T.}~\bibnamefont {Gisler}},
  \bibinfo {author} {\bibfnamefont {P.~D.}\ \bibnamefont {Kaplan}}, \bibinfo
  {author} {\bibfnamefont {A.~G.}\ \bibnamefont {Yodh}}, \ and\ \bibinfo
  {author} {\bibfnamefont {D.~A.}\ \bibnamefont {Weitz}},\ }\href@noop {}
  {\bibfield  {journal} {\bibinfo  {journal} {Phys. Rev. Lett.}\ }\textbf
  {\bibinfo {volume} {85}},\ \bibinfo {pages} {888} (\bibinfo {year}
  {2000})}\BibitemShut {NoStop}%
\bibitem [{\citenamefont {Bevington}\ and\ \citenamefont
  {Robinson}(2003)}]{Bevington:2003tc}%
  \BibitemOpen
  \bibfield  {author} {\bibinfo {author} {\bibfnamefont {P.~R.}\ \bibnamefont
  {Bevington}}\ and\ \bibinfo {author} {\bibfnamefont {D.~K.}\ \bibnamefont
  {Robinson}},\ }\href@noop {} {\emph {\bibinfo {title} {{Data Reduction and
  Error Analysis for the Physical Sciences}}}},\ \bibinfo {edition} {3rd}\ ed.\
  (\bibinfo  {publisher} {McGraw-Hill},\ \bibinfo {address} {New York},\
  \bibinfo {year} {2003})\BibitemShut {NoStop}%
\end{thebibliography}
\end{document}